\documentclass[a4paper,11pt]{article}
\usepackage{jheppub}
\usepackage{multirow}
\usepackage{amsmath,amssymb}
\usepackage{amsfonts}
\usepackage{tikzfeynman}		
\usepackage{cancel}
\usepackage{tikz}
\usepackage{mathrsfs}
\usepackage{graphicx}
\usepackage{hepunits}
\usepackage{adjustbox} 

\usepackage{colordvi,color,xcolor}
\usepackage{booktabs, colortbl}
\usepackage{multirow}
\usepackage{blindtext}
\usepackage{wasysym}
\usepackage{pifont}
\usepackage[utf8x]{inputenc}

\definecolor{gbcolor}{rgb}{.8,.3,.1}
\definecolor{gbcolor}{rgb}{.8,.3,.1}

\newcommand{\ac}[1]{#1}
\newcommand{\mc}[1]{#1}
\newcommand{\gb}[1]{#1}

\author[a]{Guillermo Ballesteros,}
\author[b]{Adrián Carmona}
\author[c]{and Mikael Chala}
\affiliation[a]{Institut de Physique Théorique, Université Paris Saclay, CEA, CNRS
91191 Gif-sur-Yvette, France}
\affiliation[b]{
Theoretical Physics Department, CERN, Geneva, Switzerland
}
\affiliation[c]{Departament de F\'isica T\`eorica, Universitat de Val\`encia and IFIC, Universitat de
Val\`encia-CSIC, Dr. Moliner 50, E-46100 Burjassot (Val\`encia), Spain}
\emailAdd{guillermo.ballesteros@cea.fr}
\emailAdd{adrian.carmona@cern.ch}
\emailAdd{mikael.chala@ific.uv.es}

\title{Exceptional Composite Dark Matter}
\abstract{
We study the dark matter phenomenology of non-minimal composite Higgs models with $SO(7)$ broken
to the exceptional group $G_2$. In addition to the Higgs, three
pseudo-Nambu-Goldstone bosons arise, one of which is electrically
neutral. A parity symmetry is enough to ensure this resonance is stable. In fact, if the breaking of the Goldstone symmetry is driven by the fermion sector, this $\mathbb{Z}_2$ symmetry is automatically unbroken in the electroweak phase. In this case, the relic
density, as well as the expected indirect, direct and collider signals
are then uniquely determined by the value of the compositeness scale, $f$.
Current experimental bounds allow to account for a large fraction of the
dark matter of the Universe if the dark matter particle is part of an electroweak triplet. The totality of the relic abundance can be accommodated if instead this particle is a composite singlet. In both cases, the scale $f$ and the dark matter mass are of the order of a few TeV.

}
\date{\today}
\begin{document}
{\flushleft CERN-TH-2017-083\\ FTUV-17-0417.9647}
\vspace{-0.5cm}
\maketitle
\section{Introduction}
 
Composite Higgs Models (CHM)~\cite{Kaplan:1983fs,Kaplan:1983sm,Dimopoulos:1981xc} are among the best motivated extensions of the Standard Model (SM) of particle physics. 
First of all, the hierarchy problem can be solved by assuming the Higgs to be a bound
state of a new strongly interacting sector. This sector is
supposed to respect an approximate global symmetry $\mathcal{G}$, which in
turn is spontaneously broken to $\mathcal{H}\subset\mathcal{G}$ at a scale $f\sim$
TeV. The
Higgs is then expected to be naturally lighter than the scale of compositeness by further assuming that it is a pseudo Nambu-Goldstone Boson
(pNGB) of this symmetry breaking pattern. Moreoever, 
this approach could also help to understand the
puzzling hierarchy of fermion masses in the SM. Indeed, the explicit breaking of
$\mathcal{G}$ by linear interactions between the SM fermions and composite operators~\cite{Kaplan:1991dc, Contino:2004vy}
translates into mixings between elementary fields and fermionic resonances at the
confinement scale $f$.
The Yukawa couplings emerge in the physical basis before electroweak (EW)
symmetry breaking, being very much dependent on the dimension of the composite operators~\cite{Panico:2015jxa}.  Therefore, making the mixing of different flavours depend on the dimension of their respective operators, their masses at the EW scale could be very different.
In particular, the top quark Yukawa coupling can be much larger than the other ones without any previous enhancement in the UV. In such a case, the explicit
breaking of the global symmetry is triggered by the top quark linear mixing.

 The requirement that one Higgs doublet is part of the pNGB spectrum restricts the amount of possible cosets. In light of this background and the ongoing tests of the scalar sector at the LHC, a systematic study of non-minimal CHMs is a timely target
that is worth aiming for.\,\footnote{The adjective \emph{non-minimal} refers to CHMs that present an extended scalar sector, contrary to the minimal model based on  $SO(5)/SO(4)$ \cite{Agashe:2004rs}. An exhaustive list of small groups that can be used to build composite Higgs models can be found, for instance, in Reference\,\cite{Bellazzini:2014yua}. According to the dimension of the global symmetry group, the smallest cosets are $SO(6)/SO(5)$, $SO(6)/[SO(4)\times U(1)]$, $SO(7)/SO(6)$ and $SO(7)/G_2$. From the point of view of the number of pNGBs, the minimal choices are $SO(6)/SO(5)$, $SO(7)/SO(6)$, $SO(8)/SO(7)$ and $SO(7)/G_2$. The first three have been already studied in the context of DM~\cite{Frigerio:2012uc,Mrazek:2011iu,Chala:2016ykx}, while $SO(8)/SO(7)$ provides a very similar phenomenology leading (only) to extra singlets.}
Thus, in this paper we consider a model based on the
symmetry breaking pattern $SO(7)/G_2$~\cite{Chala:2012af}, which gives rise to seven pNGBs transforming as
$\mathbf{7} = (\mathbf{2},\mathbf{2})+(\mathbf{3},\mathbf{1})$ under $SU(2)\times SU(2)\subset G_2$. Depending on which of the two $SU(2)$ groups is weakly gauged
(and therefore identified with the SM $SU(2)_L$) the three additional pNGBs transform as a
scalar real triplet or as three singlets of the EW group. The former constitutes a version of the inert
triplet model~\cite{FileviezPerez:2008bj} free of the hierarchy problem, in which we concentrate throughout most of the paper. 
It is worth noticing that the number of free parameters in the scalar potential of the inert triplet model is smaller than in any other extended Higgs sector (and on an equal footing with e.g.\ the singlet Higgs portal). In this sense, the coset under study is also minimal. Moreover, the collection of constraints obtained in this work will be also of relevance for triplets in other contexts.

We will highlight the main differences between the triplet and the singlet cases in the last section of the paper. In both cases, the neutral scalar can be forced
to be odd under a $\mathbb{Z}_2$ symmetry which is shown to be compatible with the strong sector
dynamics. Thus, this sector respects the symmetry $O(7) \cong SO(7) \times \mathbb{Z}_2$.\,\footnote{Note that in
what concerns the composite sector alone, the scalars are exact Goldstones and hence their
interactions are shift invariant.} The neutral extra pNGB  can then account for part or all of the
observed dark matter (DM) relic abundance, depending on its $SU(2)_L$ quantum numbers.

Both models present several
advantages in contrast to their respective elementary counterparts. Indeed, the larger symmetry on the strong sector constrains the number of independent
free parameters. If the breaking of the Goldstone symmetry is mainly driven by the fermion sector, the scalar potential depends only on three quantities, two of which can be traded by the measured values of the Higgs mass and
quartic coupling. The remaining parameter is just the compositeness scale $f$. The $\mathbb{Z}_2$ symmetry is automatically unbroken in the EW phase. On another front, the symmetry breaking pattern we consider provides a more interesting phenomenology than the
minimal  CHM of Reference\,\cite{Agashe:2004rs}. First of all, it contains
a DM candidate. Since direct and indirect DM searches bound the compositeness
scale $f$ from above, they also set a robust upper limit on the mass of the new fermionic resonances, which otherwise could not be estimated by other means than fine-tuning arguments. Moreover, since these new resonances decay into the extra scalars 
(for which there are no dedicated searches), constraints on 
vector-like fermions in light of current LHC data could therefore be weakened.

The structure of the paper is as follows. In Section\,\ref{sec:parity} we demonstrate
that the $\mathbb{Z}_2$ symmetry mentioned above can be respected by the strong sector; and we compute the pNGB
sigma model for the triplet case. There, we also discuss the representation theory for 
fermions and derive the scalar potential. The possible collider signatures are described
in Section\,\ref{sec:lhc}. The computation of the relic density,
as well as the analysis of potential direct and indirect detection signals
are presented in Section\,\ref{sec:cosmology}. Finally, in Section\,\ref{sec:singlet} we consider the possibility that the three additional scalars transform
as $SU(2)_L$ singlets rather than as a triplet, highlighting the phenomenological differences. We conclude in Section\,\ref{sec:conclusions}. Further details on the algebra of $SO(7)$ and $G_2$ are provided in Appendices\,\ref{sec:app1} and \ref{sec:app2}, while in Appendix\,\ref{sec:app3} we stress the main phenomenological consequences of sizable explicit symmetry breaking in the lepton sector.
 
\section{Viability of the $\mathbb{Z}_2$ symmetry}\label{sec:parity}

For a generic symmetry breaking pattern, let
$T^i$ and $X^a$ represent the unbroken and coset generators, respectively. Let us also define $\Pi = \Pi_a X^a$, where $\Pi_a$ runs over all pNGBs. The $d_\mu=d_{\mu}^a X^a$ symbol from the Maurer-Cartan one form
\begin{align}
	\omega_{\mu}=-i U^{-1}\partial_{\mu} U=d_{\mu}^a X^a+E_{\mu}^i T^i,\quad \mathrm{with}\quad U=\mathrm{exp}\left(i\frac{\Pi}{f}\right),
\end{align}
entering the non-linear sigma model, reads
\begin{align} \label{f22}
	d_\mu =\sum_{k=0}^{\infty}\frac{(-i)^k}{f^{k+1}(k+1)!}\mathrm{ad}_{\Pi}^k (\partial_{\mu}\Pi)_X&=\frac{1}{f}\partial_{\mu}\Pi-\frac{i}{2f^2}\left[\Pi,\partial_{\mu}\Pi\right]_X-\frac{1}{6f^3}\left[\Pi,\left[\Pi,\partial_{\mu}\Pi\right]\right]_X\nonumber\\
	&+\frac{1}{24 f^4}\left[\Pi,\left[\Pi,\left[\Pi,\partial_{\mu}\Pi\right]\right]\right]_X+\ldots
\end{align}
where we have denoted $\mathrm{ad}_{A}(B)=[A,B]$ and the subindex $X$ means the projection into the broken generators. 
It is well
known that the pNGB interactions in
symmetric spaces contain only even powers of $1/f$.
Indeed, for symmetric
cosets, $[X^a, X^b] = i f^{abi} T^i$, and hence all even powers in the expression above vanish.
Consequently, the leading-order Lagrangian in derivatives describing
the pNGB fields,
\begin{equation}
 \mathcal{L}_\sigma = \frac{1}{2}f^2 \mathrm{Tr}(d_\mu d^\mu)~,
\end{equation}
constructed out of the trace of two $d_\mu$ symbols, contains only terms with even
number of fields.

This concerns models like $SO(6)/SO(5)$~\cite{Gripaios:2009pe,Frigerio:2012uc}, $SO(7)/SO(6)$~\cite{Chala:2016ykx}, $SO(6)/SO(4)\times SO(2)$~\cite{Mrazek:2011iu} or $SU(4)\times SU(4)/SU(4)$ \cite{Ma:2017vzm} for example. We are instead interested on the coset $SO(7)/G_2$.
The corresponding generators can be found in the Appendix\,\ref{sec:app1}. They are normalized in such a way that $\text{Tr}(T^a T^b) = \delta^{ab}$ and $\text{Tr}(X^a X^b) = \delta^{ab}$. The condition $\text{Tr}(T^i X^a) = 0$ also holds. A straightforward computation shows that this space is not symmetric.
For example, $[N_1, N_2] = -i(M_3 + N_3/\sqrt{2})/\sqrt{3}$.
Nevertheles, the
leading-order sigma model still contains only even powers of $f$. This result relies on two properties. First, all
commutators with odd powers of $f$ in Equation\,\ref{f22} are parallel; likewise for all
even powers. More concretely,
\begin{equation}
 d_\mu = \frac{1}{f}\partial_{\mu}\Pi + g_1\, \hat{\Pi}^2 [\Pi,\partial_{\mu}\Pi]_X + g_2\, \hat{\Pi}^3 [\Pi,[\Pi,\partial_{\mu}\Pi]]_X~,
\end{equation}
with
\begin{equation}
	g_1 = \frac{i}{a_1}\left[-1+\cos\left(\frac{\sqrt{a_1}~\hat{\Pi}}{f}\right)\right],~
	g_2 = \frac{1}{a_2}\left[-\frac{\Pi}{f}+\frac{1}{\sqrt{a_2}}\sin\left(\frac{\sqrt{a_2}~\hat{\Pi}}{f}\right)\right].
\end{equation}
In the equations above, $a_1 = 5/6$, $a_2 = 11/18$, and
\begin{equation}
\hat{\Pi}=\sqrt{\Pi^a \Pi^a}~.
\end{equation}
Clearly, $g_1$ consists
of only even powers of $1/f$, while $g_2$ contains only odd terms. And second, one can easily check that both $ \text{Tr}(\partial_\mu\Pi [\Pi, \partial_\mu\Pi]_X)$ and $\text{Tr}([\Pi, \partial_\mu\Pi]_X [\Pi,[\Pi,\partial_{\mu}\Pi]]_X)$ vanish, and so
no odd powers of $1/f$ appear in the Lagrangian at leading order in derivatives. Given this, $SU(2)_L$ invariance implies that only terms with an even number of new multiplets are allowed, irrespectively of whether they are singlets or triplets.

We can then impose a $\mathbb{Z}_2$ symmetry under which the multiplet
containing the new neutral scalar changes sign, regardless of whether it is a singlet or a triplet. Clearly,
in light of the discussion above, this does not spoil the two-derivative
Lagrangian containing the kinetic term of the propagating fields.
Higher-order terms, instead, might be forbidden by this symmetry without
observable phenomenological consequences. As in the rest of non-minimal CHMs with DM scalars, as
well as in their renormalizable counterparts, the origin
of this symmetry is not specified and it has to be
enforced by hand. It is nonetheless interesting to prove
that this is compatible with the shift symmetry even in
a non-symmetric coset, as it is the case here.

\subsection{Gauge bosons}
Let us focus now on the triplet case.
We can compute the leading-order covariant derivative Lagrangian for the pNGBs  by promoting the derivatives in $\omega_{\mu}$ and $d_{\mu}$ to covariant derivatives, \textit{i.e.},  $\partial_{\mu}\to \partial_{\mu}-ig \sqrt{3} W_{\mu}^i M_i-ig^{\prime} B_\mu F_3$ (see Appendix\,\ref{sec:app1} for the expressions of $M_i$ and $F_3$). At lowest order in derivatives, this leads to
\begin{align}\label{eq:siglag0}
	\mathcal{L}_{\sigma}&=|D_{\mu}H|^2\left(1-\frac{1}{3f^2}|\Phi|^2\right)+\frac{1}{2}|D_{\mu}\Phi|^2\left(1-\frac{2}{3f^2}|H|^2\right)-\frac{1}{6f^2}\left[\Phi^{\dagger}t^i (D_{\mu}\Phi)\right]\left[(D^{\mu}\Phi)^{\dagger}t^i \Phi\right]\nonumber\\
	&+\frac{1}{3f^2}\partial^{\mu}(H^{\dagger}H)(\Phi^{\dagger}D_{\mu}\Phi)-\frac{2}{3f^2}|H|^2|D_{\mu}H|^2+\frac{1}{6f^2} \left[\partial_{\mu}(H^{\dagger} H)\right]^2+\mathcal{O}\left(\frac{1}{f^4}\right)~,
\end{align}
where we have defined the following $SU(2)_L\times U(1)_Y\subset SO(4)$ multiplets:
\begin{align}
	H=\frac{1}{\sqrt{2}}(h_1-i h_2,h_3+ih_4)^T\sim \mathbf{2}_{1/2}, \qquad \Phi=(\kappa_1 + i\kappa_2,-\eta,-\kappa_1-i \kappa_2)^T\sim\mathbf{3}_0,
\end{align}
and $\kappa_1,\kappa_2,\eta,h_1,h_2,h_3,h_4$ are the pNGBs associated to the broken generators $N_1, N_2, ..., N_7$. $H$ is identified with the SM-like Higgs doublet living in the $\mathbf{7}$ representation of $G_2$ and $\Phi$ with the remaining real triplet. Besides,  $t^i$, with $i=1,2,3$, read
\begin{align}
	t^1=\frac{1}{\sqrt{2}}\left(
\begin{array}{ccc}
 0 & 1 & 0 \\
1& 0 & 1 \\
 0 & 1 & 0 \\
\end{array}
	\right),\quad t^2=\frac{1}{\sqrt{2}}\left(
\begin{array}{ccc}
 0 & -i & 0 \\
 i& 0 & -i \\
 0 & i& 0 \\
\end{array}
\right),\quad t^3=\left(
\begin{array}{ccc}
 1 & 0 & 0 \\
 0 & 0 & 0 \\
 0 & 0 & -1 \\
\end{array}
\right).
	\label{eq:gen1}
\end{align}
We have also redefined $f\to -f/(2\sqrt{2/3})$. We will keep this convention henceforth. We also identify $h = h_3$ with the physical Higgs boson.
The part of the above Lagrangian involving only $h$ can be easily summed to all orders in $1/f^2$, resulting in
\begin{align}\label{eq:siglag}
	\mathcal{L}_{\sigma}\supset \frac{1}{2}(\partial_{\mu}h)^2+\left[\frac{1}{4}g^2f^2\sin^2\bigg(\frac{h}{f}\bigg)W_{\mu}^{+}W^{\mu-}+\frac{1}{8}(g^2+g^{\prime 2})f^2\sin^2\bigg(\frac{h}{f}\bigg) Z_{\mu}^{+}Z^{\mu-}\right]~,
\end{align}
where we have defined
\begin{align}
	Z_{\mu}=\cos\theta_W W_{\mu}^3-\sin\theta_W B_{\mu}, \qquad A_{\mu}=\sin\theta_W W_{\mu}^3+\cos\theta_W B_{\mu}
\end{align}
and $\tan\theta_W=g^{\prime}/g$ as usual. In particular, we can see that after EW symmetry breaking (EWSB), the $W$ and $Z$ bosons get masses
\begin{align}
	m_W^2=\frac{1}{4}g^2f^2\sin^2\bigg(\frac{\langle h \rangle}{f}\bigg), \qquad m_Z^2=\frac{1}{4}(g^2+g^{\prime 2})f^2\sin^2\bigg(\frac{\langle h \rangle}{f}\bigg)~,
\end{align}
with $\langle h \rangle$ the Higgs vacuum expectation value (VEV), which differs from the SM EW VEV $v = f\sin{(\langle h \rangle/f)}\sim 246$ GeV.
It is also clear that $\rho=m_W^2/m_Z^2c_W^2=1$,
as expected due to the custodial symmetry $SO(4)\subset G_2$. The ratio of the tree level coupling between the Higgs and the massive gauge bosons to the corresponding SM coupling differs from unity by the amount:
\begin{equation} \label{eq:hvv}
 R_{hVV} = \sqrt{1-\xi}, \qquad \xi = \frac{v^2}{f^2}.
\end{equation}
Clearly, given that $f\sim {\rm TeV}$, the ratio $\xi \ll 1$. If the SM group $SU(2)_L\times U(1)_Y$ of $SO(7)$ is the only gauged group in the EW sector, the global symmetry $SO(7)$ is broken explicitly. This becomes manifest in the (non-vanishing) scalar potential. In order to compute it, we promote
the $SU(2)_L\times U(1)_Y$ gauge bosons to be in the adjoint of the global $SO(7)$ with the help of \emph{spurion} fields. For this aim, let us order the generators of $SO(7)$ as $T^{\hat{a}}=\{F^1,\ldots,F^7,M^1,\ldots,M^7,N^1,\ldots,N^7\}$. We can then write
\begin{align}
	A_{\mu}^{\hat{a}}=W_{\mu}^i\Xi^{i\hat{a}}+B_{\mu}\Upsilon^{\hat{a}},\qquad \hat{a} = 1,...,21, \qquad i = 1, 2, 3~.
\end{align}
The spurions $\Xi^{i\hat{a}}$ and $\Upsilon^{\hat{a}}$ are given explicitly by the expressions:
\begin{align}
	\Xi^{i\hat{a}}=\sqrt{3}~\delta^{(i+7)\hat{a}},\qquad \Upsilon^{\hat{a}}=\delta^{\hat{a}3}~.
\end{align}
Formally, they also transform in the $\mathbf{21}$ of $SO(7)$. The \emph{dressed} field $A_{\mu}^D=U^{-1}A_{\mu}^{\hat{a}} T^{\hat{a}}U$ transforms under $g\in SO(7)$ as $h(\Pi,g)A_{\mu}^D h^{-1}(\Pi,g)$, with $h\in G_2$, and  decomposes as a sum of irreps of $G_2$. The same happens to the dressed spurions $\Xi^i_D=U^{-1}\Xi^{i\hat{a}}T^{\hat{a}}U$ and $\Upsilon_D=U^{-1}\Upsilon^{\hat{a}} T^{\hat{a}} U$, with the difference that the index $i$ spans an $SU(2)_L$ triplet.

The gauge contribution to the scalar potential consists therefore of the different invariants that can be built out of the $G_2$ irreps within $\Xi^i_D$ and $\Upsilon_D$ and can be expressed as an expansion in powers of $g/g_{\rho}$ and $g^{\prime}/g_{\rho}$, with $g_{\rho}$ the characteristic coupling of the strong sector vector resonances.    Taking into account that  $\Xi^i_D$ and $\Upsilon_D$  decompose as $\mathbf{7}\oplus \mathbf{14}$ under $G_2$, we obtain only one independent invariant at leading order:
\begin{align}\label{eq:gaugepot}
	\ac{V_{\rm gauge}(\Pi)=\frac{3}{4}\frac{m_{\rho}^4}{(4\pi)^2}\left(\frac{g}{g_{\rho}}\right)^2\frac{1}{\hat{\Pi}^2}\left[\left(6\tilde{c}_1 +2\tilde{c}_2\frac{g^{\prime 2}}{g^2}\right)|H|^2+8\tilde{c}_1|\Phi|^2\right]\sin^2\left(\frac{\hat{\Pi}}{f}\right),}
\end{align}
where \ac{$\tilde{c}_{1,2}$  are $\lesssim 1$ dimensionless numbers, $m_{\rho}\sim g_{\rho} f$ is the typical mass of the vector resonances}, and we have used naive dimensional analysis~\cite{Giudice:2007fh,Panico:2012uw,Chala:2017sjk} to account for the $\hbar$ and mass dependence of the radiative potential.

\subsection{Fermions}\label{sec:fermions}

The mixing between the elementary fermions and the composite sector breaks explicitly the global symmetry $SO(7)$, because the former transform in complete representations of  the EW subgroup only. Let us first focus on the quark sector. The mixing Lagrangian can be written as:
 \begin{align}
	 \mathcal{L}_{\rm mix}\sim \lambda_{q}^{ij} \bar{q}_{\alpha L}^{i} (\Delta_{q}^\alpha)^I (\mathcal{O}_{q}^j)_I + \lambda_{u}^{ij} \bar{u}_R^{i} (\Delta_{u})^I (\mathcal{O}_{u}^j)_I + \lambda_{d}^{ij} \bar{d}_R^{i} (\Delta_{d})^I (\mathcal{O}_{d}^j)_I + \textrm{h.c.}~.
	 \label{eq:mix}
 \end{align}
The indices $i,j = 1,2,3$ run over the three quak generations. $\alpha = 1,2$ and $I$ are instead $SU(2)_L$ and $SO(7)$ indices, respectively. The couplings $\lambda^q_{33}$ and $\lambda^u_{33}$ are supposed to be order one and much larger than all other couplings. This is expected from the dependence of the quark Yukawas on these numbers, namely $y_{u,d}^{ij} \sim \lambda^{\dagger ik}_q\lambda_{u,d}^{kj}/g_*$ where $g_{\ast}$ is the typical coupling of the strong sector fermionic resonances. 
The spurion fields $\Delta$ are incomplete multiplets of $SO(7)\times U(1)_X$.\,\footnote{The addition of the extra (unbroken) $U(1)_X$ is necessary to accommodate the SM fermion hypercharges.} Formally, they transform in the same representations as the corresponding composite operators, $\mathcal{O}$. We assume that the third generation right and left quarks mix with composite operators transforming in the $\mathbf{1}_{2/3}$ and the $\mathbf{35}_{2/3}$ of $SO(7)\times U(1)_X$, respectively. This is motivated by the following branching rules under $G_2\times U(1)_X$ and the EW gauge group:
 \begin{align}
	 \mathbf{35} = \mathbf{1}_{2/3}\oplus \mathbf{7}_{2/3}\oplus \mathbf{27}_{2/3} = \mathbf{1}_{2/3}\oplus \mathbf{2}_{\pm 1/2}\oplus\mathbf{3}_0\oplus \mathbf{1}_{2/3}\oplus\mathbf{2}_{\pm 1/2}\oplus ...	  \label{eq:spurions}
 \end{align}
 where the ellipsis stands for higher-dimensional representations in the branching rule of the $\mathbf{27}$.\,\footnote{Under $SO(4)\cong SU(2)_L\times SU(2)_R\subset G_2$\,: 
	$ \mathbf{7}=(\mathbf{2},\mathbf{2})\oplus (\mathbf{3},\mathbf{1})$ and  $\mathbf{27}=(\mathbf{1},\mathbf{1})\oplus (\mathbf{2},\mathbf{2}) \oplus (\mathbf{3},\mathbf{3})\oplus (\mathbf{4},\mathbf{2}) \oplus (\mathbf{5},\mathbf{1}).$}
 In order not to break the EW symmetry, the spurions $\Delta_q^\alpha$ can only have non-zero entries in the doublets. However, the $\mathbb{Z}_2$ symmetry requires the components along the second one to vanish.\,\footnote{Let $(i,j)$ run over the non-vanishing entries of the spurions $\Delta_q^{1}$ and $\Delta_q^{2}$ (see Appendix\,\ref{sec:app1}). Then, note that under the $\mathbb{Z}_2$, the elements $(i,j)$ of the $U$ matrix do not change sign. Therefore, the spurions  are even eigenstates:  $\mathbb{Z}_2(\Delta_q^1) = \Delta_q^1$ and $\mathbb{Z}_2(\Delta_q^2) = \Delta_q^2$. On the contrary, the spurion accomodating the second doublet includes, for example, a non-vanishing entry in $(1,4)$, while $U_{14} \sim h_1\kappa_1$, that changes sign.} Its explicit expression can be found in Appendix\,\ref{sec:app1}.
Similarly to the gauge boson case, the dressed spurion $\Delta_{qD}^{\alpha}=U^{-1}\Delta_{q}^\alpha U$ transforms under $g\in SO(7)$ as $\Delta^\alpha_{qD}\to h(\Pi,g)\Delta_{qD}^\alpha h^{-1}(\Pi,g)$ with $ h\in G_2$ and decomposes as a sum of irreps: $\Delta_{qD}^\alpha=\bigoplus_m \Delta_{q}^{\alpha m}$. The fermion contribution to the scalar potential can be written as
\begin{align}
	V_{\rm fermion}(\Pi)\approx  m_{\ast}^4  \frac{N_c}{(4\pi)^2}\left[\left(\frac{|\lambda_{q}|}{g_{\ast}}\right)^2\sum_j c_j V_{j }(\Pi)+\left(\frac{|\lambda_{q}|}{g_{\ast}}\right)^4\sum_k c_{k}^{\prime}  V_{k}^{\prime}(\Pi)+\ldots\right],
\end{align}
where \ac{$m_{\ast}\sim g_{\ast} f$ is the typical mass scale of the fermionic resonances} and again we have used naive dimensional analysis  to estimate the parametric dependence of the potential, with $N_c=3$ and $c_i$ and $c_i^{\prime}$ order one dimensionless numbers. $V_{j}(\Pi)$, $V_{k}^{\prime}(\Pi)$ and the terms indicated by the ellipsis are the different invariants that can be built out of two, four and higher number of insertions of $\Delta_{q I}^m$, respectively. For simplicity, we have also defined $\lambda_q=\lambda_q^{33}$. Note that no terms proportional to $\lambda_u$ appear, because the right-top mixing, being a full singlet, does not break the global symmetry.

The scalar potential consists then of the left-handed top-induced and the gauged-induced potentials. However, the latter can be neglected if
\begin{align}
	\ac{\tilde{c}_1 g^2 g_{\rho}^2\lesssim 2\pi^2 \lambda_H\Rightarrow \tilde{c}_1g_{\rho}^2 \lesssim 8, }
\end{align}
where $\lambda_H\sim 0.13$ is the SM Higgs quartic coupling and we have disregarded the hypecharge contribution for simplicity.
Indeed, if this inequality holds, all observables computed taking into account only the top-induced potential are (almost) unaffected when the gauge potential is also included. This ocurrs, in particular, for $\tilde{c}_1\sim 0.1$ and moderately large values of $g_{\rho}$; and also if $\tilde{c}_1\sim 1$ and $g_{\rho}\sim 1$. Whereas the former possibility may involve some additional tuning, the latter arises naturally at large values of $f$, which, as we will see, are the ones preferred to account for the observed  DM relic abundance. We consider this scenario hereafter. Then,
\begin{align}
	V(\Pi)\approx m_{\ast}^2 f^2 \frac{N_c}{16\pi^2}y_t^2 \left[ c_1 V_1(\Pi)+c_2 V_2(\Pi)\right],
	\label{eq:pot}
\end{align}
where $|\lambda_q|$ has been traded by the top Yukawa coupling $y_t$ (see below) and we have defined
\begin{align}
	\sum_{\alpha}	\left|(\Delta_{qD}^{\alpha})_{88}\right|^2 \sim V_1(\Pi)&=\frac{|H|^2}{\hat{\Pi}^2}\sin^2\left(\frac{2\hat{\Pi}}{f}\right), \\
	\sum_{\alpha}\sum_{i=1}^7\left|(\Delta_{qD}^{\alpha})_{i8}\right|^2 \sim V_2(\Pi)&=\frac{|H|^2}{\hat{\Pi}^2}\cos\left(\frac{4\hat{\Pi}}{f}\right)+\frac{3|H|^2+2|\Phi|^2}{\hat{\Pi}^2}\cos\left(\frac{2\hat{\Pi}}{f}\right).
\end{align}
The scalar potential above depends only on two independent unknowns, $c_1$ and $c_2$. They parametrize the two invariants constructed out of $ \mathbf{1}\times \mathbf{1}$ and $\mathbf{7}\times \mathbf{7}$ in Equation\,\ref{eq:spurions}, respectively. 
Note that the potential features only  an even number of powers of $\Phi$. This is actually true \mc{at} any order in $\lambda_q/g_*$, \mc{because} the spurions are $\mathbb{Z}_2$--even and the $\mathbb{Z}_2$ invariance of the potential requires $\Phi$ to  appear always squared. Let us further keep the leading-order potential in the expansion in powers of $1/f^2$. This can be matched to the renormalizable piece
\begin{align}
	V_{\rm renorm}(H,\Phi)=\mu^2_{H} |H|^2+\lambda_H |H|^4+\frac{1}{2}\mu_{\Phi}^2 |\Phi|^2+\frac{1}{4}\lambda_{\Phi}|\Phi|^4+\lambda_{H\Phi}|H|^2|\Phi|^2~.
	\label{eq:ren}
\end{align}
The five parameters in Equation\,\ref{eq:ren} can be expressed in terms of the parameters $c_1$ and $c_2$. These can be traded by the measured values of the SM EW VEV and the Higgs quartic coupling, $\lambda_H$. Up to the scale $f$, all parameters of phenomenological relevance are then \textit{predictions}. These are given in Table~\ref{tab:pot}. It can be checked that $\langle\Phi\rangle = 0$ in the EW phase, since $\mu_\Phi^2 > 0$ and $\lambda_{H\Phi} > 0$. And so, as anticipated, the $\mathbb{Z}_2$ symmetry is not spontaneously broken. We would like to point out that the negative sign of $\lambda_\Phi$ does not necessarily imply a (potentially dangerous) runaway behaviour at high energies, where the effective description we use fails. The existence or not of such a behaviour, and of a possible minimum at higher energies would depend on the specific way that the model is completed in the UV.

It is also worth stressing that, after EWSB, the masses of the charged and neutral components of $\Phi$ are both equal to
%
\begin{align}\label{eq:DMmass}
	m_{\Phi}^2=\mu_{\Phi}^2+v^2\lambda_{H\Phi}=\frac{2}{3}f^2\lambda_{H}\left[1-\frac{9}{4}\frac{v^2}{f^2}\right]+...~,
\end{align}
where the ellipsis stands for terms that are further suppressed by powers of $v^2/f^2$. 
The splitting between the masses of the charged components and that of the neutral one comes only from (subdominant) radiative EW corrections. It can be estimated to be $\Delta M\sim 166$ MeV~\cite{Cirelli:2005uq}.

Finally, we can also compute the top Yukawa Lagrangian:
\begin{align}\label{eq:topyukawa}
	\sum_{\alpha}\bar{q}_L^{\alpha}\left(\Delta_{qD}^{\alpha}\right)^{\dagger}_{88}t_R+\mathrm{h.c.}\sim \mathcal{L}_{\rm yuk}&= c_t \lambda_q \left(\bar{q}_L\tilde{H} t_R\right)\frac{f}{\hat{\Pi}} \sin\left(\frac{2\hat{\Pi}}{f}\right)+\mathrm{h.c.}\nonumber\\
	&= -y_t (\overline{q}_L \tilde{H} t_R)\bigg[1-\frac{2}{3\mc{f^2}}\Phi^2+...\bigg]+\mathrm{h.c.}\,,
\end{align}
where
$c_t$ is an order one dimensionless parameter encoding the UV dynamics and the product $-2 c_t \lambda_q$  has been traded by the top Yukawa, $y_t$. This Lagrangian is explicitly $\mathbb{Z}_2$-invariant. 
If we \mc{add} all terms involving only the Higgs boson, the ratio of the tree level coupling of the Higgs
to the massive top quark to the corresponding SM coupling is:
\begin{equation}\label{eq:htt}
 R_{htt} = \frac{1-2\xi}{\sqrt{1-\xi}}, \qquad \xi = \frac{v^2}{f^2}.
\end{equation}
\begin{table}[t]
	\begin{center}
	\begin{adjustbox}{width=\textwidth}
		\begin{tabular}{lcccc}
			\toprule
			Parameter&$\mu_{H}^2$&$\mu_{\Phi}^2$&$\lambda_{\Phi}$&$\lambda_{H\Phi}$\\
			\midrule
			Value&$-v^2\lambda_H$&$\frac{2}{3}f^2\lambda_H\left(1-\frac{8}{3}\frac{v^2}{f^2}\right)$&$-\frac{4}{9}\lambda_H\left(1-\frac{8}{3}\frac{v^2}{f^2}\right)$&$\frac{5}{18}\lambda_{H}\left(1+\frac{32}{15}\frac{v^2}{f^2}\right)$\\
						\bottomrule
		\end{tabular}
		\end{adjustbox}
		\caption{Values of the different parameters of the renormalizable scalar potential as a function of $f$, to order $\mathcal{O}(v^2/f^2)$. $v\sim 246$ GeV and $\lambda_H\sim 0.13$ stand for the \mc{SM} EW VEV and the Higgs quartic coupling, respectively.}
		\label{tab:pot}
	\end{center}

\end{table}

\section{Collider signatures}~\label{sec:lhc}
Different collider searches bound $f$ from below. Among the
more constraining ones, we find monojet analyses, searches
for disappearing tracks, measurements of the Higgs to
diphoton rate \mc{and EW precision tests}. The small probability for an emission of a
hard jet in association with two invisible $\Phi$ particles
makes monojet searches less efficient than the other two.

The Higgs decay width into photons is modified by order $\xi$
due to the non-linearities of the Higgs couplings, as stated
in Eqs.~\ref{eq:hvv} and~\ref{eq:htt},
and, to a smaller extent, due to the new charged scalars
that can run in this loop-induced process. The width (taking
into account both effects) is 
given by~\cite{Carena:2012xa,Espinosa:2012qj}:
\begin{equation}
\Gamma(h\rightarrow \gamma\gamma) = \frac{\alpha^2 v^2 m_h^3}{1024 \pi^3} \bigg[\frac{g^2}{2 m_W^2} \sqrt{1-\xi} A_1(\tau_W) + \frac{4 y_t^2}{3 m_t^2} \frac{1-2\xi}{\sqrt{1-\xi}} A_{1/2}(\tau_t) + \frac{\lambda_{H\Phi}}{m_\kappa^2} A_0(\tau_\kappa)
\bigg]^2~
\end{equation}
where $\tau_i = 4 m_i^2/m_h^2$, $A_0(x) = -x^2(x^{-1} - F(x^{-1}))$, $A_{1/2}(x) = 2x^2(x^{-1} + (x^{-1} - 1)F(x^{-1}))$ and $A_1(x) = -x^2(2x^{-2} + 3x^{-1} + 3(2x^{-1} -1)F(x^{-1}))$, while the function $F$ is given by $F(x) = \arcsin^2{\sqrt{x}}$.
The Higgs production cross section via gluon fusion is also modified by order
$\xi$ effects:
\begin{equation}
\sigma(gg\rightarrow h) = \frac{(1-2\xi)^2}{1-\xi}\sigma^\text{SM}(gg\rightarrow h)~,
\end{equation}
with $\sigma^\text{SM}$ the SM production cross section.
Given that $\xi > 0$  and $\lambda_{H\Phi} > 0$ (see Table \ref{tab:pot}), the production cross section
times branching ratio is always smaller than in the SM.
A combination of 7 and 8 TeV data from both ATLAS and CMS~\cite{Khachatryan:2016vau}
sets a lower bound of $0.66$ on $\sigma(gg\rightarrow h \rightarrow\gamma\gamma)/\sigma_{\text{SM}}(gg\rightarrow h \rightarrow\gamma\gamma)$ at 95\,\% C.L.
This translates into a bound on $f \gtrsim 800$ GeV. \mc{EW precision tests~\cite{Ghosh:2015wiz} push this bound to $f\gtrsim 900$ GeV}. Searches at future colliders (see for example Reference\,\cite{Klute:2013cx}) would determine the Higgs to diphoton cross section with a much better accuracy. In particular, the region $f\lesssim 1.5$ TeV is expected to be probed in Higgs searches at future facilities.

Finally, searches for dissappearing tracks are sensitive to
pair-production and the subsequent decay of the new
scalars. Indeed, the small splitting between the charged and
the neutral components of $\Phi$ implies that the former has
a decay length exceeding a few centimeters. This produces tracks
in the tracking system that have no more than a few associated
hits in the outer region, in contrast with most of the SM processes.
To our knowledge, the most
constraining search of this kind was performed by the ATLAS
Collaboration in~\cite{Aad:2013yna} (similar results were found in the
CMS analysis of Reference\,\cite{CMS:2014gxa}). Searches
of this type using 13 TeV data are not yet published.

The ATLAS search is optimized for a Wino
(\textit{i.e.} a generic triplet fermion, $\chi$) with a width of $\sim 160$ MeV,
corresponding to a lifetime of $\sim 0.2$ ns, whose charged
components therefore decay predominantly into the neutral
one and a soft pion. In this respect, the search applies equally
well to our scalar triplet. The search rules out any mass
below $\sim 270$ GeV, corresponding to a
production cross section of $\sim 0.25$ pb. The latter takes
into account the production of all $\chi^+\chi^-$, $\chi^+\chi^0$ and $\chi^-\chi^0$. The corresponding
bound on $f$ is therefore given by the value at
which the production cross section in the scalar case equals
the previous number (note that, for the same mass, the scalar
and triplet cross sections can be very different).

In order to compute this cross section at the same level of
accuracy as the one considered in the experimental reference,
\textit{i.e.} at NLO in QCD, we first implement the
renormalizable part of our model in \texttt{Feynrules v2}~\cite{Alloul:2013bka}. UV
and $R_2$ terms~\cite{Ossola:2008xq} are subsequently computed by means of
\texttt{NLCOT}~\cite{Degrande:2014vpa}. The interactions are then exported to an
\texttt{UFO} model that is finally imported in \texttt{MadGraph v5}~\cite{Alwall:2014hca}
to generate parton-level events from which the total cross
section is computed for all values of $f$ in the range
$500, 600, ..., 1500$ GeV.
The bound on $f$ turns out to be only $f \gtrsim 650$ GeV.
However, future facilities could easily exceed the reach of Higgs searches \cite{Cirelli:2014dsa, Low:2014cba, Mahbubani:2017gjh}.
For example, a naive reinterpretation of the results in Reference\,\cite{Cirelli:2014dsa} 
suggests that values of $f$ as large as $\sim 3.5$ TeV could
be tested in a future 100 TeV $pp$
collider.

\section{Searches for dark matter}\label{sec:cosmology}

In this section of the paper we analyze the extent to which $\eta$, the neutral component of the scalar triplet $\Phi$ of our model, can contribute to the DM of the Universe, given the current experimental constraints. As we anticipated in the Introduction, the compatibility of a global $\mathbb{Z}_2$ symmetry with the breaking pattern $SO(7)/G_2$ allows to forbid $\eta$ decays. This neutral particle, which couples to the SM through weak interactions and does not couple directly to the photon is, a priori, a good weakly interacting massive particle (WIMP) DM candidate with the adequate mass scale. Remarkably, the mass of $\eta$ and its relic abundance are, in our case, entirely determined by the scale $f$, which makes the model extremely predictive. In this last respect, the model is on pair with other simple implementations of the WIMP idea, such as the Minimal DM model \cite{Cirelli:2005uq}. 

We recall that the total annihilation rate, $\langle \sigma\, v\rangle$, and the relic abundance, $\Omega\,h^2$, of any thermal relic  satisfy the approximate relation
 \begin{align} \label{approx}
\Omega\,h^2\sim \frac{3\times 10^{-27}}{ \langle \sigma\, v\rangle}\, {\rm cm}\, {\rm s}^{-1}\,,
 \end{align}
 where the brackets indicate the average over the thermal velocity distribution. Thus, if a thermal relic explains the totality of the DM abundance ($[\Omega\, h^2]_{\rm DM}\sim 0.11$\,\cite{Ade:2013zuv}), it must have an annihilation rate of the order of  ${ \langle \sigma\, v\rangle}\sim 3 \times 10^{-26}\, {\rm cm}\, {\rm s}^{-1}$. 
 
\begin{figure}[t]
 \begin{center}
  \includegraphics[width=0.8\columnwidth]{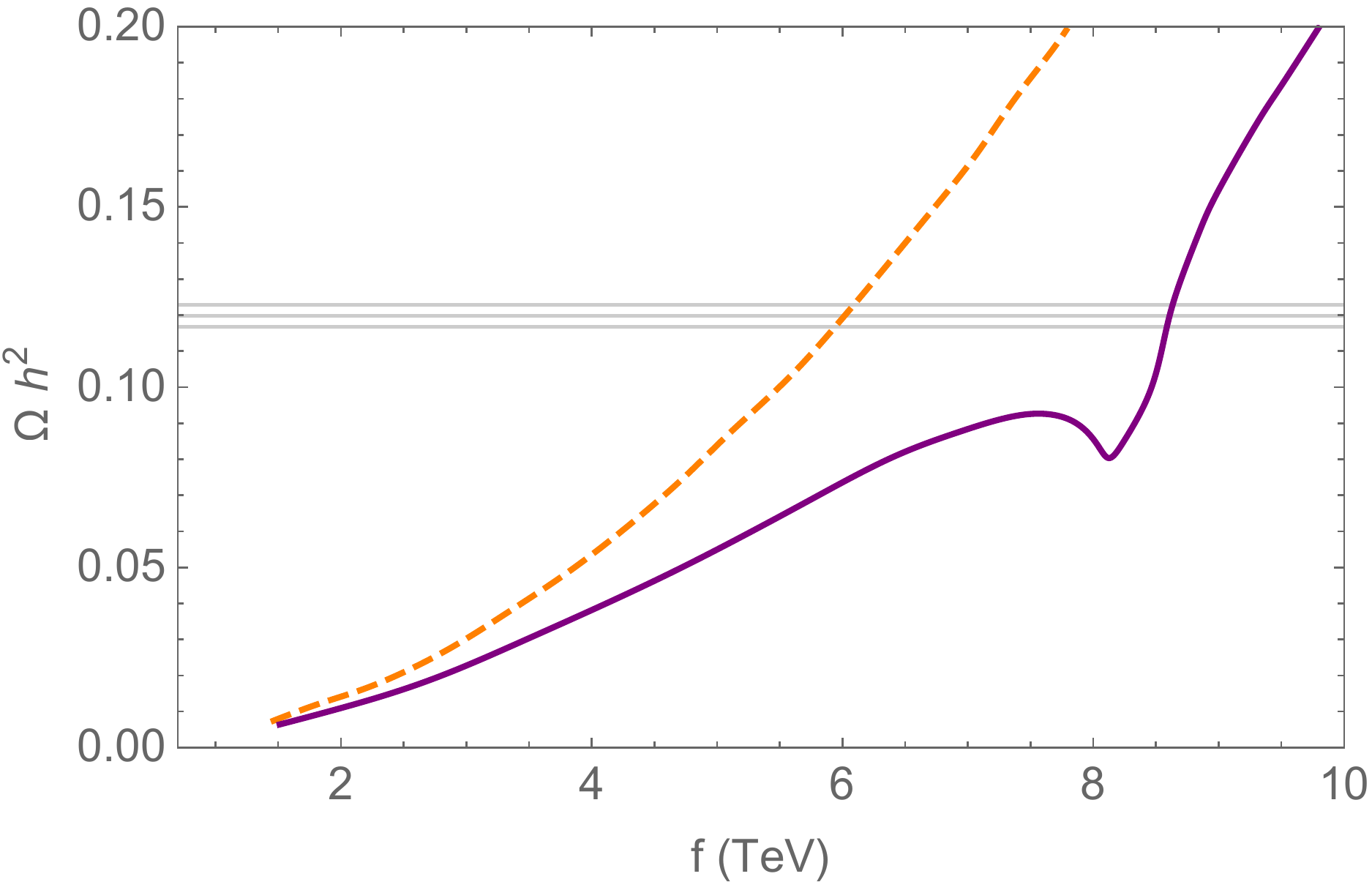}
 \end{center}
 \caption{The dependence of the relic abundance $\Omega\, h^2$ of $\eta$ as a function of the compositeness scale $f$, with (continuous) and without (dashed) \mc{non-perturbative} effects; see \cite{Cirelli:2007xd}. The horizontal lines show the measured central value and a 95\,\% C.L.\ interval around it as determined by Planck \cite{Ade:2013zuv}.}
 \label{fig:Omegavsf}
\end{figure}

Given the expression \ref{approx}, the relic abundance turns to be roughly proportional to the mass squared of the thermal relic. The relic abundance for the neutral component of a scalar triplet as a function of its mass was computed in \cite{Cirelli:2007xd}. Including \mc{non-perturbative effects}, it was found that a mass of $\sim 2.5$ TeV is required to obtain the measured DM abundance.  In Figure~\ref{fig:Omegavsf}, we recast this result as a function of the compositeness scale $f$ of our model, which is related to the mass of the neutral component of the triplet, $\eta$, through Equation\,\ref{eq:DMmass}.{\footnote{The result shown in Figure~\ref{fig:Omegavsf} assumes that the portal coupling coupling $\lambda_{H\Phi}$ is negligible. This is indeed the case in our model, since $\lambda_{H\Phi}$ is roughly an order of magnitude smaller than the gauge couplings.}}
As shown in Figure~\ref{fig:Omegavsf}, a scale $f\sim 8.6$~TeV is required in this case to account for the totality of the DM in the Universe. In the remaining of this section we explore whether this scale is compatible with the current bounds from the LHC and direct and indirect detection experiments; and we determine how much DM can be accounted for by $\eta$. 

\subsection{Direct detection}
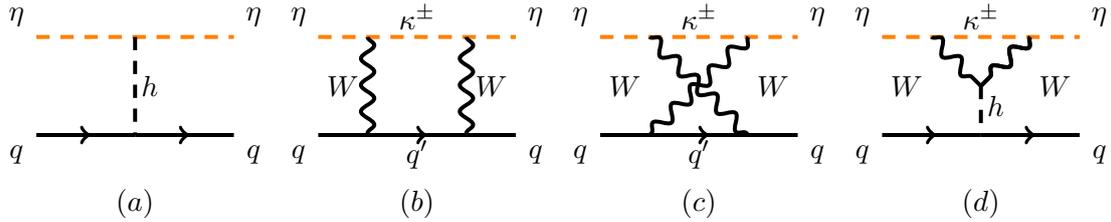
\begin{figure}[!t]
	\begin{center}
	\begin{tikzpicture}[line width=1.5 pt, scale=.65]
		\draw [color=orange, dash pattern=on 5pt off 4pt]		(-2,1.0) -- (2.,1.);
		\draw [fermion]		(-2,-1) -- (0,-1);
		\draw [fermion]		(0,-1) -- (2.,-1);
		\draw [dash pattern=on 5pt off 4pt]		(0,1) -- (0,-1);
			\node at (-2.4, 1.4) {$\eta$};
			\node at (2.4, 1.4) {$\eta$};
			\node at (0.3, 0) {$h$};
			\node at (-2.4,-1.4) {$q$};
			\node at (2.4,-1.4) {$q$};
			\node at (0, 1.4) {$\phantom{\kappa^{\pm}}$};
			\node at (0, -1.4) {$\phantom{q^{\prime}}$};
			\node at (0, -2.4) {$(a)$};
	\end{tikzpicture}
	\begin{tikzpicture}[line width=1.5 pt, scale=.65]
		\draw [color=orange, dash pattern=on 5pt off 4pt]		(-2,1.0) -- (2.,1.);
		\draw [fermion]		(-2,-1) -- (2,-1);
		\draw [vector]		(-1,1) -- (-1,-1);
		\draw [vector]		(1,1) -- (1,-1);
			\node at (-2.4, 1.4) {$\eta$};
			\node at (2.4, 1.4) {$\eta$};
			\node at (0, 1.4) {$\kappa^{\pm}$};
			\node at (0, -1.4) {$q^{\prime}$};
			\node at (-1.5, 0) {$W$};
			\node at (1.5, 0) {$W$};
			\node at (-2.4,-1.4) {$q$};
			\node at (2.4,-1.4) {$q$};
			\node at (0, -2.4) {$(b)$};
	\end{tikzpicture}
	\begin{tikzpicture}[line width=1.5 pt, scale=.65]
		\draw [color=orange, dash pattern=on 5pt off 4pt]		(-2,1.0) -- (2.,1.);
		\draw [fermion]		(-2,-1) -- (2,-1);
		\draw [vector]		(-1,1) -- (1,-1);
		\draw [vector]		(1,1) -- (-1,-1);
			\node at (-2.4, 1.4) {$\eta$};
			\node at (2.4, 1.4) {$\eta$};
			\node at (0, 1.4) {$\kappa^{\pm}$};
			\node at (0, -1.4) {$q^{\prime}$};
			\node at (-1.5, 0) {$W$};
			\node at (1.5, 0) {$W$};
			\node at (-2.4,-1.4) {$q$};
			\node at (2.4,-1.4) {$q$};
			\node at (0, -2.4) {$(c)$};
	\end{tikzpicture}
	\begin{tikzpicture}[line width=1.5 pt, scale=.65]
		\draw [color=orange, dash pattern=on 5pt off 4pt]		(-2,1.0) -- (2.,1.);
		\draw [vector]		(-1,1) -- (0,0);
		\draw [vector]		(1,1) -- (0,0);
		\draw [vector]		(1,1) -- (0,0);
		\draw [dash pattern=on 5pt off 4pt]		(0,0) -- (0,-1);
		\draw [fermion]		(-2,-1) -- (0,-1);
		\draw [fermion]		(0,-1) -- (2.,-1);
			\node at (-2.4, 1.4) {$\eta$};
			\node at (2.4, 1.4) {$\eta$};
			\node at (0, 1.4) {$\kappa^{\pm}$};
			\node at (-1.5, 0) {$W$};
			\node at (1.5, 0) {$W$};
			\node at (-2.4,-1.4) {$q$};
			\node at (2.4,-1.4) {$q$};
			\node at (0.3, -0.4) {$h$};
			\node at (0, -2.4) {$(d)$};
	\end{tikzpicture}
		\caption{$(a)$: Tree level contribution to the cross section for  spin-independent scattering of DM on nucleons. $(b)$-$(d)$: Representative examples of loop-induced diagrams which are relevant for the $m_{\Phi}$--independent piece of the cross section. }

		\label{fig:all}
\end{center}
\end{figure}

The cross section  for spin-independent scattering of DM on nucleons  has a tree-level  contribution proportional to the portal coupling $\lambda_{H\Phi}$ \cite{Cline:2013gha}: 
\begin{align}
	\sigma(\eta N\to \eta N)_{\rm tree}&= \frac{ \lambda_{H\Phi}^2  m_{N}^4f_N^2}{ \pi m_h^4 m_{\Phi}^2}	\approx \frac{25  m_N^4f_N^2}{864 \pi f^2 v^4 \lambda_{H}}\left(1+\frac{391}{60}\frac{v^2}{f^2}\right)~,
\end{align}
 which scales  with the inverse of the DM mass squared  and arises  from the tree level exchange of a Higgs boson on $t$-channel (see Figure~\ref{fig:all}~$(a)$), 
where
\begin{align}
	f_N=\sum_{q}f_q=\sum_q\frac{m_q}{m_N}\langle N| \bar{q}q| N\rangle=0.30\pm 0.03~,
\end{align}
 and $m_N=\frac{1}{2}(m_n+m_p)\sim 1\,$GeV is the nucleon mass. However, due to the presence of derivative interactions $\sim i\Phi \overleftrightarrow{\partial_{\mu}}\Phi W^{\mu}$ in the Lagrangian, there is also a loop induced contribution independent of  $m_{\Phi}$. It comes from the virtual exchange of $W$ bosons (which is insensitive to $\lambda_{H\Phi}$), because they bring down a $p^2\approx m_{\Phi}^2$  term which precisely cancels the $1/m_{\Phi}^2$ factor  coming from the phase space integral, see \textit{e.g.} the diagrams in Figures~\ref{fig:all}~$(b)$-$(d)$.  Such cross section has been computed in the heavy WIMP effective theory (HWET) \cite{Hill:2013hoa, Hill:2014yka, Hill:2014yxa, Solon:2016yai}. The leading term in the $1/m_{\Phi}$ expansion (valid therefore for  $m_{\Phi}\gg m_W\gg m_q$) reads
 \begin{align}
	 \sigma(\eta N\to \eta N)_{\rm HWET}=1.3^{+0.4+0.4}_{-0.5-0.3}\times 10^{-2}\, \mathrm{zb}~.
 \end{align}
 This value includes  contributions from two-loop diagrams and is \emph{universal}, in the sense that it only depends on the $SU(2)_L$ quantum numbers of the heavy particle, while further details of the model (such as the spin of the WIMP or its possible interaction with the Higgs) enter only through $1/m_{\Phi}$ corrections. 
 
\begin{figure}[t]
	\begin{center}
	\includegraphics[width=0.8\columnwidth]{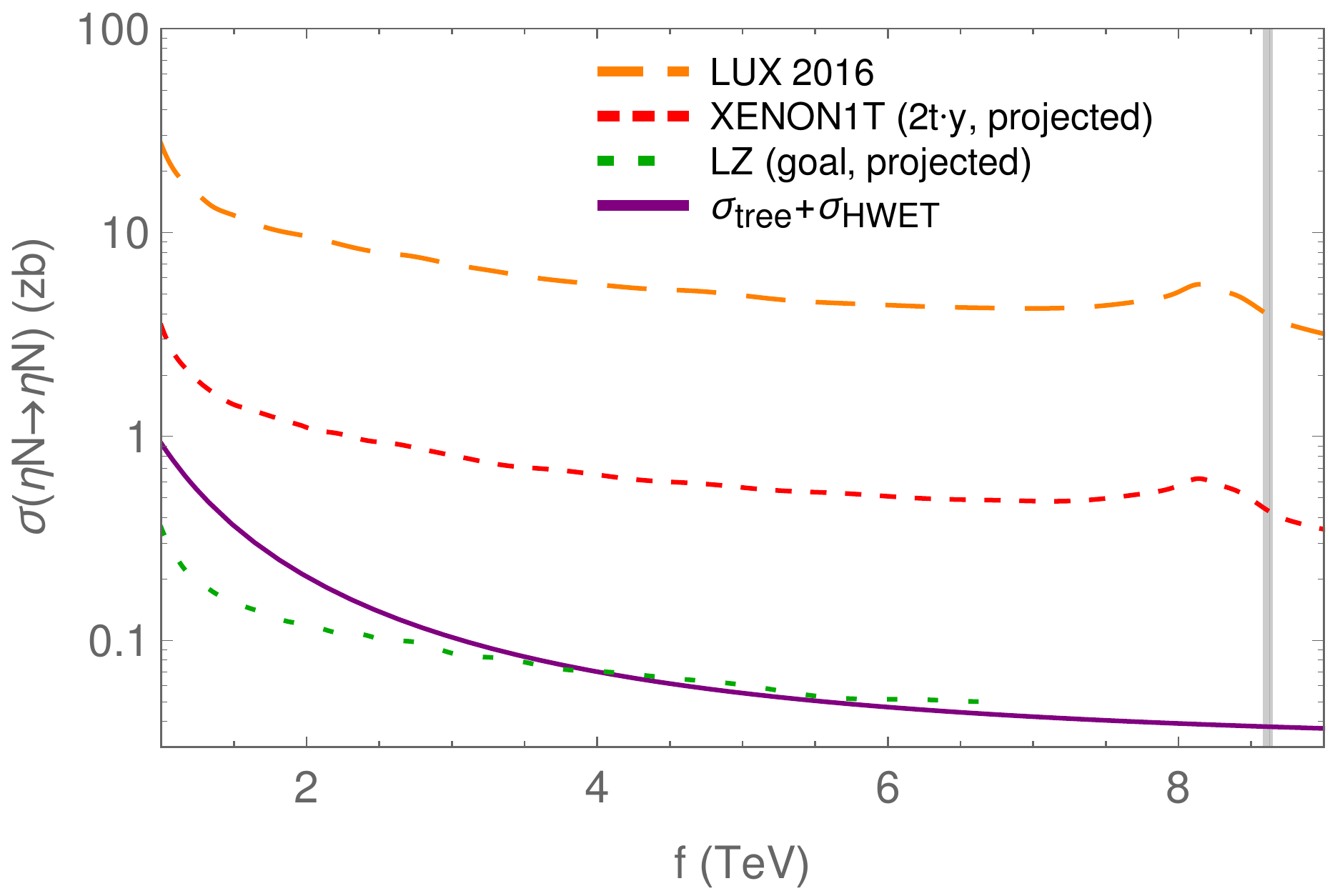}
		\caption{Estimate for the spin-independent direct detection cross section $\sigma(\eta N \to \eta N)=\sigma(\eta N\to \eta N)_{\rm tree}+ \sigma(\eta N\to \eta N)_{\rm HWET}$ as a function of the compositeness scale $f$ (purple) versus the current limits \ac{(linearly rescaled with the DM abundance)}  from LUX (dashed orange) \cite{Akerib:2016vxi} and the projected  exclusion limits at $95$\,\% C.L. for LZ (dotted green) \cite{Szydagis:2016few} and XENON1T of 2 years in 1 ton (dashed red)  \cite{Aprile:2015uzo}. }
		\label{fig:dd}
	\end{center}
\end{figure}

In order to provide a conservative estimate of the sensitivity of current and projected direct detection experiments to this model, we show in Figure~\ref{fig:dd} the sum of both contributions to the spin-independent cross section as a function of the compositeness scale (purple) versus the latest limits from LUX \cite{Akerib:2016vxi} (dashed orange)  together with the projected sensitivities for LZ (dashed green) \cite{Szydagis:2016few} and XENON1T (dashed red) \cite{Aprile:2015uzo}. The latter are properly rescaled by $[\Omega h^2]_{\rm DM}/\Omega h^2$, \gb{which} takes into account that $\eta$ could be \gb{just} a subcomponent of the whole relic abundance. In order to be more conservative, we have used the 1$\sigma$ upper values for both contributions in $\sigma(\eta N\to \eta N)$. It should be noted that this is only an estimate of the DM-nucleon cross section, since the validity of the HWET breaks down for low values of $f$ (and hence low masses). We also neglect possible interference effects. The low sensitivity of current experiments ensures that making more accurate predictions is not needed. Interestingly, the order of magnitude of the estimated cross section is in the \gb{ballpark} of the aimed sensitivity for LZ, making the model accessible via direct detection in the near future.

\subsection{Indirect detection}
Indirect DM searches use astrophysical and cosmological observations to look for the effects of SM particles into which DM is assumed to decay or annihilate. Concretely, they focus on the Cosmic Microwave Background (CMB) and the detection of gamma  and cosmic rays originating from decays or annihilations of DM. 
We consider the effects due to the $W$ boson pairs that are produced in the annihilation of $\eta$ particles, which is \gb{the main} relevant channel in our case. We restrict our attention to three different kinds of indirect probes, which provide the current most constraining bounds on the annihilation rate of WIMPs: the CMB, gamma rays \gb{coming} from dwarf spheroidal galaxies and gamma rays from the center of the Milky Way. The last of these two observables have specific intrinsic uncertainties due to our limited knowledge about the DM distribution inside galaxy halos; and therefore the CMB leads to a more robust bound. It is worth stressing that quite generically DM constraints coming from indirect detection experiments are less robust than direct detection ones and even less than those coming from colliders, such as the LHC. Again, the reason is the required modelling of astrophysical phenomena in indirect detection experiments. Nevertheless, when taken at face value these constraints are the most stringent ones on our model and therefore they deserve to be considered in depth. However, we warn the reader that the percentages of the DM relic density that we derive in what follows should be taken as an indicative approximation.

\begin{figure}[t]
 \begin{center}
  \includegraphics[width=0.8\columnwidth]{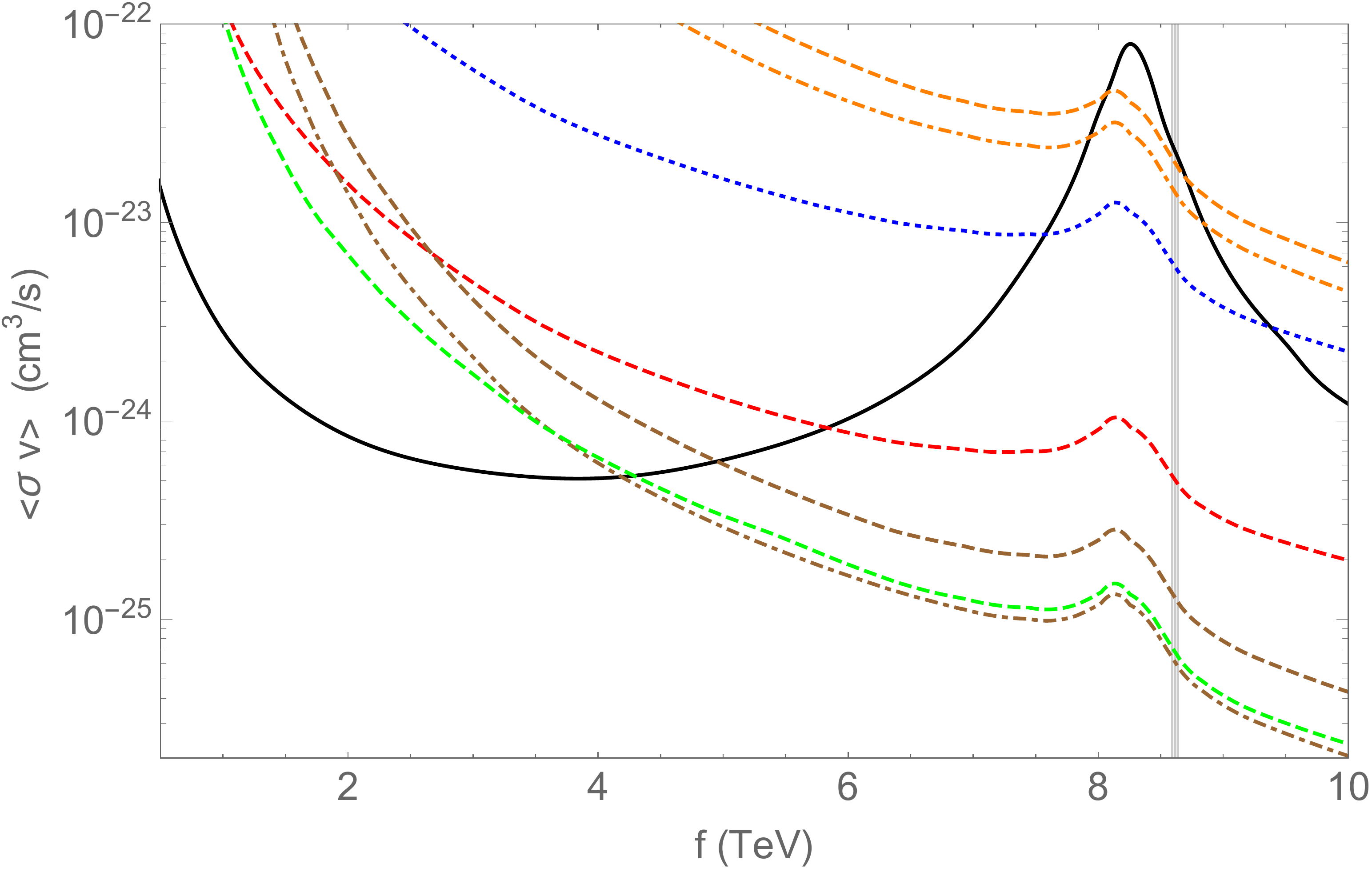}
 \end{center}
	\caption{The thermally averaged annihilation rate of $\eta\eta$ into $W^+W^-$ as a function of the compositeness scale, $f$. The black continuous line is the theoretical prediction of the model. From top to bottom we also show the upper bounds \ac{(rescaled with the square of the DM abundance)} from the following observations: H.E.S.S.\ dwarf spheroidal galaxies [Burkert (dashed) and NFW (dot-dashed) profiles, in orange], CMB from Planck (blue, dotted), the combination of FERMI and MAGIC dwarf spheroidal galaxies (red, dashed), and the Milky Way center as seen by H.E.S.S.\ [NFW (dashed) and Einasto (dot-dashed) profiles, in brown]. See the main text for references. The green dashed line is the expected sensitivity of CTA for observations of the Milky Way center assuming an Einasto profile \cite{Silverwood:2014yza}.  The vertical band at $f\simeq 8.5$ TeV locates the scale that gives the total DM abundance with a 95\,\% C.L.\ from the prediction of the model; see also Figure~\ref{fig:Omegavsf}.}
 \label{ID}
\end{figure}

In Figure~\ref{ID} we compare the theoretical prediction for the annihilation rate $\langle \sigma\, v \rangle$  of $\eta$ particles from \cite{Cirelli:2007xd},\,\footnote{See Figure~3 in Reference\,\cite{Cirelli:2007xd} for a scalar triplet.}  as a function of the scale $f$,  with the current bounds from Planck, H.E.S.S.\ and FERMI+MAGIC. The shape of all the curves, peaking around $\sim$ 8.2 TeV is due to the use of Figure~\ref{fig:Omegavsf} to rescale the bounds on $\langle \sigma\, v \rangle$ from the various collaborations. This is needed to account for the fact that the event rate of any annihilation process scales as the square of the local density of annihilating particles, which in the case of WIMPs can be assumed to be approximately proportional to the relic density $\Omega\, h^2$. The usual indirect detection upper bounds on the DM annihilation rate assume that all the DM in the Universe corresponds to a single WIMP species of a given mass. In order to include the possibility that $\eta$ explains only a fraction of the total DM abundance, the experimental bounds have thus to be multiplied by a factor $([\Omega\, h^2]_{\rm DM}/[\Omega\, h^2])^2\simeq 0.012\, [\Omega\, h^2]^{-2}$. Obviously, this takes into account the dependence of the abundance $\Omega\, h^2$ of $\eta$ on its mass (or, equivalently, on the scale $f$),  as shown in Figure~\ref{fig:Omegavsf}. 

The weakest indirect detection bounds that we consider come from the observation by the Cherenkov radiation telescope H.E.S.S.\  of dwarf spheroidal galaxies, which are strongly DM dominated systems and supposed to be free from other gamma ray emission. These bounds correspond to the two upper lines of Figure~\ref{ID}, see \cite{Abramowski:2014tra}. The distance between them comes from their different assumptions for the radial distribution of DM in those galaxies. The upper curve assumes a Burkert profile \cite{Burkert:1995yz}, which features a constant inner density core, whereas the lower one is for an NFW profile \cite{Navarro:1995iw,Navarro:1996gj}, which peaks at the center. A more stringent limit from dwarf spheroidals is reported by the collaborations of the FERMI satellite and the Cherenkov telescope MAGIC. The combination of their respective data leads to the red dashed curve \cite{Ahnen:2016qkx}, assuming an NFW profile.  Although these data are more constraining, a direct comparison to the results of H.E.S.S.\ is not straightforward, since the details of the assumed profiles are different. 

Notwithstanding the importance of dwarf spheroidal galaxies for indirect DM detection, the center of the Milky Way is thought to be the strongest gamma ray emitter and therefore an important candidate for a potential indirect DM detection. Clearly, the choice of DM profile is critical for the interpretation of these observations, but unfortunately the DM distribution in the center of our galaxy is uncertain.  Moreover, there is a number of baryonic astrophysical sources of gamma rays which need to be accounted for when considering the possible emission from the Galactic center. These backgrounds are not well known either and this implies a large source of uncertainty in addition to the choice of DM profile. The pair of brown lines at the bottom of Figure~\ref{ID} represent the current constrains from the Milky Way galactic center obtained by H.E.S.S.\ \cite{2016PhRvL.117k1301A} for two profile choices: NFW (dashed) and Einasto \cite{1965TrAlm...5...87E} (dot-dashed). Although these observations appear to be the most stringent, it has to be emphasized that they are also  the ones whose interpretation carries a larger uncertainty. It is interesting to point out that modifying the parameters of the NFW profile these limits can be weakened slightly above the Einasto curve, see \cite{2016PhRvL.117k1301A}.

As we mentioned earlier, the most robust bounds come from the CMB, and in particular from the Planck satellite. The blue dashed line of Figure~\ref{ID} represents the bound obtained in the analysis of \cite{2016PhRvD..93b3527S}. According to the CMB upper bound on the annihilation rate, our exceptional DM model can account for as much as 80\,\% of the DM abundance of the Universe at 95\,\% C.L., as can be read from Figures~\ref{ID} and \ref{fig:Omegavsf}. This requires a value of the composite scale $f\simeq 7.5$ TeV, which corresponds to a triplet mass of $\sim 2.2$ TeV. If instead we take the strongest limits from the Galactic center from H.E.S.S.\, the maximum percentage of the DM abundance that can be explained by $\eta$ is at most 36\,\%, corresponding to $f\simeq 4.25$~TeV and a mass of $\sim 1.25$ TeV. Given that indirect detection sets the strongest upper bounds in our model, we can conclude that a significant amount of the DM of the Universe might be in the form of the neutral component of our triplet. 

Future indirect detection data could in principle test the model with improved sensitivity in the range of $f$ that is relevant for DM. The Cherenkov \gb{Telescope Array} CTA \cite{2013APh....43..189D}, which should start taking data by 2021, may currently be the best proposal that could contribute to that goal.  Several CTA sensitivity estimates exist in the literature, in particular, for DM annihilation in the Galactic center into a pair of $W$ bosons \cite{Wood:2013taa,Silverwood:2014yza, Lefranc:2015pza}. \gb{These estimates vary depending on the assumptions made about the final configuration design of the telescope array, the observational strategy (including its timespan) and several other factors. In Figure \ref{ID} we report the forecast of reference \cite{Silverwood:2014yza} for DM annihilation into $W$ bosons, appropriately rescaled with the DM abundance. According to \cite{Silverwood:2014yza}, it appears that once systematics effects are accounted for, the upper bound that will be reachable with CTA for this specific channel might not be too dissimilar from the most stringent current limits obtained by H.E.S.S.\ \cite{2016PhRvL.117k1301A}. However, the value of the cross section that will be attainable with CTA for the range of masses that interests us is estimated to be a factor $\sim 4.5$ lower in \cite{Lefranc:2015pza}; but this number accounts only for statistical errors. It is clear that a proper comparison between current bounds and different forecasts would require, at the very least, the use of the same DM profile.

In principle, the model could also be constrained from searches of monochromatic gamma lines due to the annihilation of DM into two photons  in the central regions of the Milky Way. To the best of our knowledge, the latest and most stringent upper bounds on the cross section for this process in the relevant range of mass have been obtained by the H.E.S.S.\ collaboration \cite{Abramowski:2013ax,Abdalla:2016olq}. A DM mass of $\sim 1.2$ TeV approximately corresponds to $f\sim 4$ TeV, which is the scale at which the H.E.S.S.\ limits on DM annihilation into $W^+W^-$ intersect the theoretical prediction; see Figure \ref{ID}. The current strongest bound for $\eta\eta\rightarrow \gamma\gamma$ and DM masses around that value is $\langle \sigma v\rangle \sim 10^{-27}$ cm$^3/{\rm s}$ at 95\% C.L., assuming an Einasto profile. For a scalar triplet with zero hypercharge, this cross section has been computed (including the Sommerfeld effect) in \cite{Cirelli:2007xd}. After the adequate rescaling with the DM abundance, the theoretical prediction is $\langle \sigma v\rangle \sim 5\times 10^{-28}$ cm$^3/{\rm s}$, which is an order of magnitude lower than the aforementioned upper bound. Although, once more, the DM profile dependence is an important source of uncertainty,\footnote{See {\it e.g.} References \cite{Cirelli:2015bda, Garcia-Cely:2015dda}.}  this channel is not more constraining in our case than $\eta\eta\rightarrow W^+W^-$. 

A  CTA sensitivity estimate applicable for $\eta\eta\rightarrow \gamma\gamma$ was produced in \cite{Ovanesyan:2014fwa} under the assumption of an NFW profile. Translating this estimate to the relevant range of $f$ and after rescaling by the DM abundance, it gives $\langle \sigma v\rangle \sim 3\times 10^{-29}$ cm$^3/{\rm s}$, which is an order of magnitude lower than the theoretical estimate. This means that CTA observations of monochromatic gamma lines should allow to probe the model beyond the current H.E.S.S.\ bound from $\eta\eta\rightarrow W^+W^-$. This type of search may in fact be able to test all the range of $f$ for which the limits on the annihilation cross section into $W$ bosons still allows to account for a significant fraction of the DM relic abundance.}

\section{Singlet case}\label{sec:singlet}
The model we have explored so far provides a hyperchargeless scalar triplet as  DM candidate. This is a consequence of weakly gauging one particular $SU(2)$ of the two global ones respected by the strong sector, which makes the fundamental representation of $SO(7)$ decompose as $\mathbf{7}=\mathbf{2}_{1/2}\oplus\mathbf{3}_0$ under the EW subgroup $SU(2)_L$. However, as it was done in \cite{Chala:2012af}, one can also weakly gauge the other $SU(2)$ within $SO(4)\cong SU(2)_L\times SU(2)_R$, under which  $\mathbf{7}=\mathbf{2}_{1/2}\oplus \mathbf{1}_0\oplus\mathbf{1}_{\pm 1}$, obtaining an isospin singlet as potential DM candidate. We follow this path in this section, highlighting the specific differences between the two cases. 

\paragraph{Gauge contribution to the scalar potential}
Contrary to the triplet case, the potential of the new charged (and hypercharged) scalars receives contributions proportional to $g^\prime$. Equation\,\ref{eq:gaugepot} has to be modified by:
\begin{align}
	\ac{V_{\rm gauge}(\Pi)=\frac{3}{4} \frac{m_{\rho}^4}{(4\pi)^2}\left(\frac{g}{g_{\rho}}\right)^2\frac{1}{\hat{\Pi}^2}\left[\left(6\tilde{c}_1 +2\tilde{c}_2\frac{g^{\prime 2}}{g^2}\right)|H|^2+8\tilde{c}_2 \frac{g^{\prime 2}}{g^2} \kappa^{+}\kappa^-\right]\sin^2\left(\frac{\hat{\Pi}}{f}\right)~.}
\end{align}
This term modifies the mass splitting between $\kappa^\pm$ and $\eta$, with respect to the case of the triplet. It gives a contribution $(m_{\kappa^\pm}-m_\eta)/m_{\eta}\sim {g'}^2/(N_c\, y_t^2)\sim 0.05$.
\paragraph{Fermion contribution to the scalar potential}
In the singlet case, the charged and neutral scalars do not exchange gauge bosons, and hence the first cannot decay into the second. In order to avoid an over-abundance of stable charged particles, for which stringent constraints exist \cite{Gould:1989gw,DeRujula:1989fe,Dimopoulos:1989hk} new sources of sizable explicit symmetry breaking have to be considered. Being the second heaviest fermion, we assume that this effect is driven by the bottom quark. There are many different possible embeddings of the right-handed bottom quark, $b_R$. However, not all of them respect the $\mathbb{Z}_2$ symmetry $\eta\leftrightarrow  -\eta$ that makes the singlet scalar stable or generate a bottom Yukawa coupling at leading order in $\lambda_q \lambda_{b_R}$, with $\lambda_{b_R}=\lambda_{d}^{33}$. We consider the case where $b_R$ mixes with the $\mathbf{7}_{2/3}$ within the $\mathbf{21}_{2/3}$, since it fulfills both conditions. Then, Equation\,\ref{eq:pot} still holds, but it has to be supplemented by the (sub-leading) bottom contribution to the scalar potential:
\begin{align}
	V_{\rm bottom}(\Pi)=\frac{N_c}{(4\pi)^2} m_{\ast}^4 \hat{c}_1\left(\frac{|\lambda_{b_R}|}{g_{\ast}}\right)^2\frac{1}{\hat{\Pi}^2}\left[2|H|^2+\eta^2+\kappa^{+}\kappa^-\right]\sin^2\left(\frac{\hat{\Pi}}{f}\right)~.
\end{align}
This term of the potential also contributes to breaking the mass degeneracy between $\kappa^\pm$ and $\eta$, giving $(m_{\kappa^\pm}-m_\eta)/m_{\eta}\sim (g_*\, y_b)^2/y_t^4\sim 6\times 10^{-4}\,g_*\lesssim 10^{-2}$. In addition, the following Yukawa couplings are generated
\begin{align}
	&	\sum_{\alpha}\sum_{i=1}^7\bar{q}_L^{\alpha}\left(\Delta_{qD}^{\alpha}\right)^{\dagger}_{i8}\left(\Delta_{b}\right)_{i8}b_R+\mathrm{h.c.}\sim\mathcal{L}_{\rm yuk,b}= \frac{c_b}{2\sqrt{6}} \frac{\lambda_q^{\ast} \lambda_{b_R}}{g_{\ast}}\frac{f}{\hat{\Pi}}\sin\left(\frac{\hat{\Pi}}{f}\right)\bar{q}_L\left[H\cos\left(\frac{\hat{\Pi}}{f}\right)\right.\nonumber\\
	&\left.- i\tilde{H}\frac{3}{\sqrt{2}}\frac{\kappa^{+}}{\hat{\Pi}}\sin\left(\frac{\hat{\Pi}}{f}\right)\right] b_R +\mathrm{h.c.} =- y_b \bar{q}_L\left[H - i\tilde{H}\frac{3}{\sqrt{2}}\frac{\kappa^{+}}{f}\ldots\right]b_R+\mathrm{h.c.}\,,
\end{align}
where $c_b$ is a dimensionless order one parameter and we have traded $\frac{c_b}{2\sqrt{6}} \frac{\lambda_q^{\ast} \lambda_{b_R}}{g_{\ast}}$ by $-y_b$ in the second expression. This provides a vertex $i\frac{3 }{\sqrt{2}}\frac{m_b}{f} \bar{t}_L  b_R \kappa^{+}$ that makes $\kappa^\pm$ decay into $tb$. 

\paragraph{Collider implications}
Searches for disappearing tracks do not constrain the singlet case. So, measurements of the Higgs couplings dominate the reach of current and future facilities. On another hand, analyses of invisible Higgs decays~\cite{Aad:2015txa,Khachatryan:2016whc} forbid only $f \lesssim 300$ GeV.  Monojet searches are further suppressed by the small coupling of the Higgs boson to $\eta$. This can be also produced in gluon fusion via loops of top quarks, but its coupling to the latter is suppressed with respect to the top Yukawa by order $\xi$. The charged scalar can be instead produced via gauge interactions. However, the small rate together with the unclean final state containing tops and bottoms, make its discovery challenging at the LHC. Future facilities could probe this channel, though.

\begin{figure}[t]
 \begin{center}
  \includegraphics[width=0.8\columnwidth]{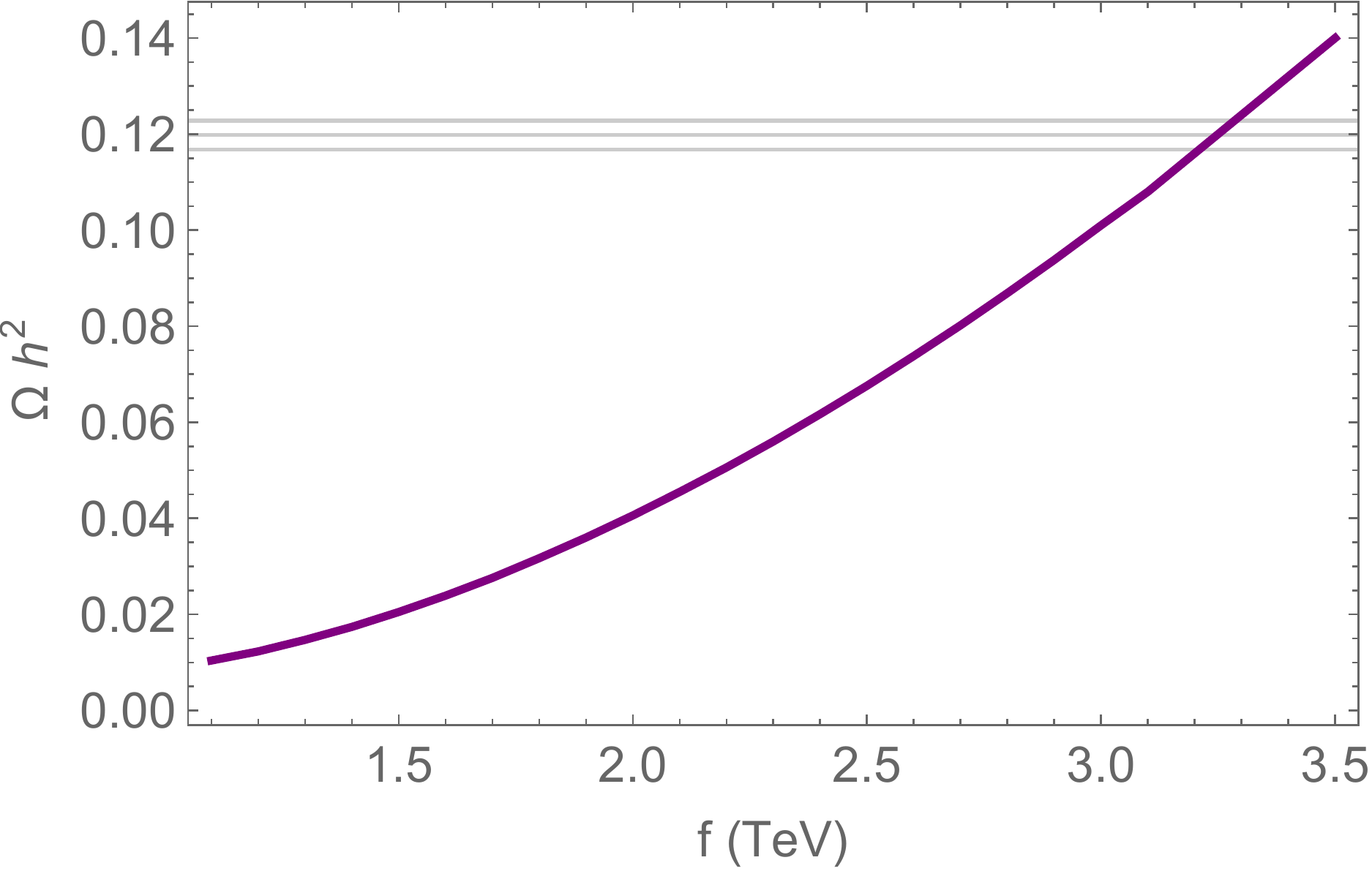}
 \end{center}
 \caption{The dependence of the relic abundance $\Omega\, h^2$ in the singlet case as a function of the compositeness scale $f$. The horizontal lines show the measured central value and a 95\,\% C.L.\ interval around it as determined by Planck \cite{Ade:2013zuv}.}
 \label{fig:singlet}
\end{figure}

\paragraph{Relic density}
Given the small splitting between the masses of the charged and the neutral components (which is driven by the small gauge induced potential), the DM particles are not expected to annihilate into $\kappa^+\kappa^-$ final states. As a consequence, the main annihilation channels are $t\overline{t}$ as well as $W^+W^-, ZZ$ and $hh$. The first channel dominates for small $f\lesssim 1.7$ TeV; see Figure~\ref{fig:singletx}, right panel. The main reason is that the annihilation into tops proceeds \mc{also} via contact interactions \mc{(analogous to the ones coming from  Equation~\ref{eq:topyukawa})}, suppressed by $1/f^2$. In the unitary gauge, other DM interactions, instead, are driven by the Higgs portal. This receives contributions from both the scalar quartic coupling in the potential $\lambda_{H\Phi}$ and from derivative operators like $|H|^2(\partial^\mu\eta)^2$, appearing in the sigma-model Lagrangian. The ratio between these two is given by (see Table\,\ref{tab:pot})
\begin{equation}
\frac{1}{2}\lambda_{H\Phi} \frac{f^2}{m_\eta^2}\sim \frac{1}{2} \frac{5}{18}\lambda_H \frac{3 f^2}{2\lambda_H f^2} \sim 0.2~,
\end{equation}
and therefore the derivative interactions dominate. The main annihilation channel for large values of $f$ is $\eta\eta\rightarrow W^+W^-$, as shown in Figure~\ref{fig:singletx}.\,\footnote{For an exhaustive discussion of the effects of higher-dimensional operators in related models see for example References\,\cite{Fonseca:2015gva, Bruggisser:2016nzw,Bruggisser:2016ixa}.}

\mc{Non-perturbative effects, like the Sommerfeld enhancement of the formation of bound states are not relevant}. For each value of $f$ we have computed the relic density by just using \texttt{micrOMEGAs v3}~\cite{Belanger:2010pz}. The result is shown in Figure~\ref{fig:singlet}, alongside the current observational band (as in Figure~\ref{fig:Omegavsf}). It turns out that the whole relic abundance can be explained by this model with $f\sim 3$ TeV, for which $m_\eta \sim 900$ GeV. As we will see, current direct and indirect searches do not exclude this possibility. However, future experiments will have the required sensitivity to test this prediction.

\paragraph{Direct searches}
\begin{figure}[t]
 \begin{center}
    \includegraphics[width=0.8\columnwidth]{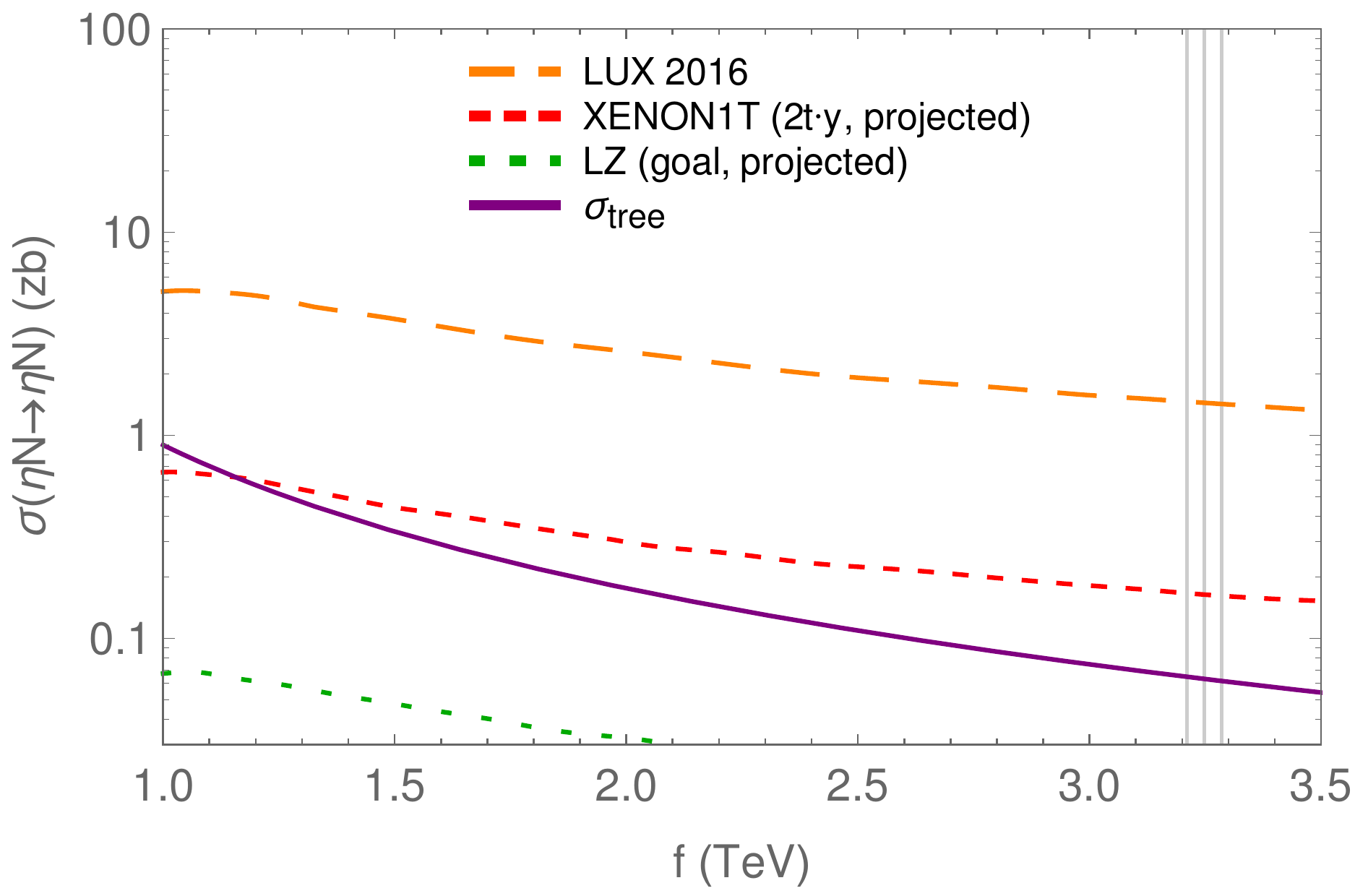}
       \end{center}
	\caption{Spin-independent direct detection cross section as a function of the compositeness scale, $f$, in the singlet DM case. We show the theoretical estimate as a purple continuous line. We also show the current limits \ac{(linearly rescaled with the DM abundance)} from LUX (dashed orange) \cite{Akerib:2016vxi} and the projected  exclusion limits at $95$\,\% C.L. for LZ (dashed green) \cite{Szydagis:2016few} and XENON1T of 2 years in 1 ton (dashed red) \cite{Aprile:2015uzo}.}
 \label{fig:singletd}
\end{figure}

Contrary to the triplet case, the DM-nucleon interaction proceeds only via the Higgs exchange. As it can be seen in Figure~\ref{fig:singletd}, current searches are not constraining enough for this model, but future experiments will definitely probe the whole parameter space. 

\paragraph{Indirect searches}

The total thermally averaged cross section for DM annihilation as a function of $f$ is shown as a  black continuous line in the left panel of Figure~\ref{fig:singletx}. As we already mentioned, DM particles annihilate mostly into $W^+W^-$ for sufficiently large values of the compositeness scale, see Figure~\ref{fig:singletx}, right panel. For this reason, we also show in the left panel the current upper constraints on $\eta\eta\rightarrow W^+W^-$ from observations of the Galactic center by H.E.S.S.\ (brown dot-dashed line) \cite{2016PhRvL.117k1301A}. This bound assumes that DM particles annihilate exclusively into $W^+W^-$ (and, as we already discussed, it is the most stringent one for this kind of process). We show as well an estimate of the future sensitivity of CTA for the same process (green dashed line) \cite{Silverwood:2014yza}. The remarks we made in the triplet case concerning this estimate and its comparison to the results of  \cite{2016PhRvL.117k1301A} also apply now. Clearly, the prediction of the singlet model for the total cross section appears to be well below the current bound and the future sensitivity for the dominant channel. We conclude that the singlet variant of exceptional composite DM is viable for all the interesting values of $f$.  In particular, for $f\sim 3.25$~TeV, all the DM abundance in the Universe can be accounted for. A substantial improvement in the sensitivity of the next generation of indirect probes will be needed to test this result.

\begin{figure}[t]
 \begin{center}
    \includegraphics[width=0.52\columnwidth]{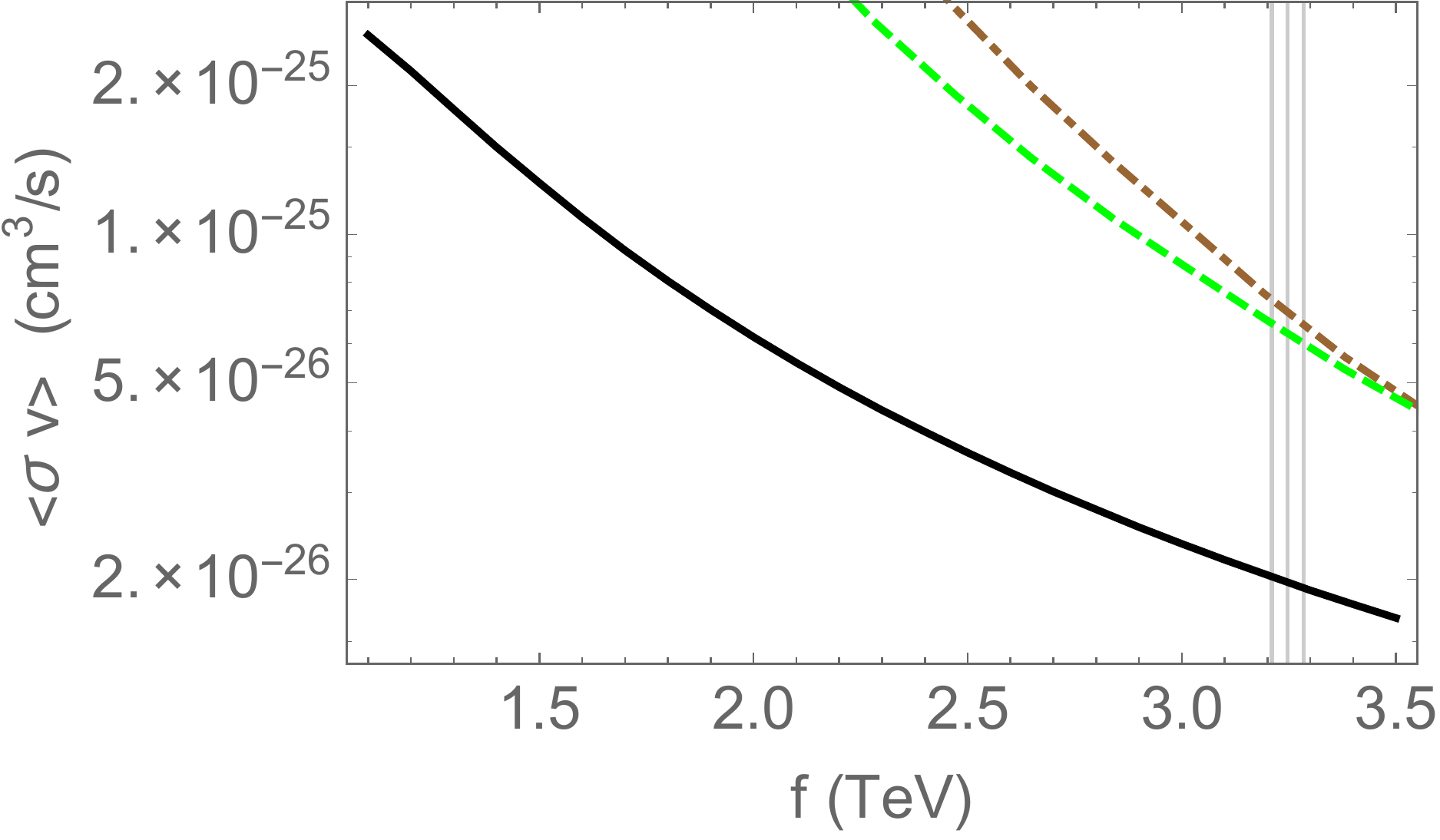}
      \includegraphics[width=0.47\columnwidth]{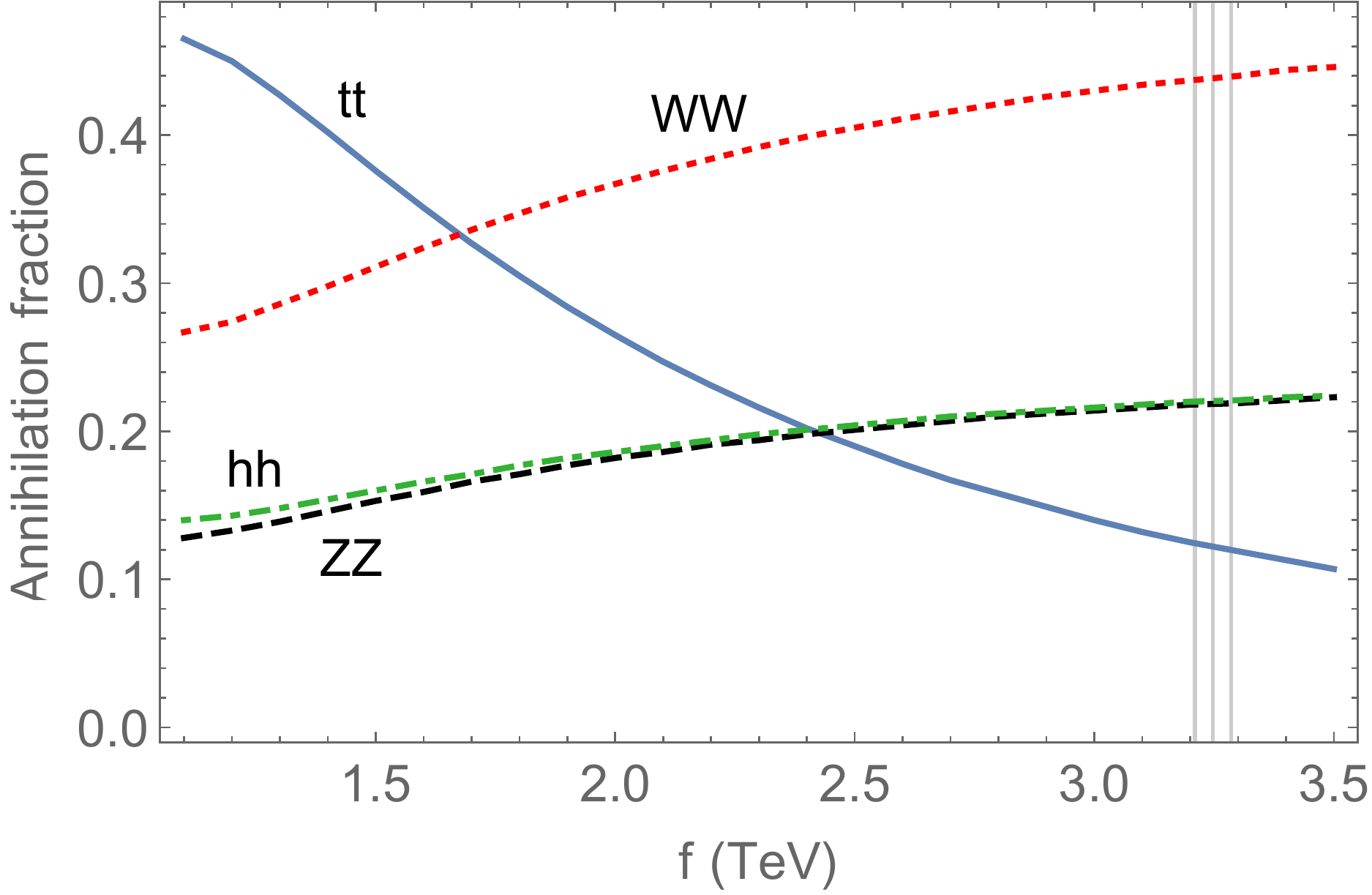}
       \end{center}
	\caption{Left panel: The theoretical prediction for the total thermally averaged annihilation rate of DM particles as a function of the compositeness scale, $f$, for the singlet DM case (black continuous line). We also show the current upper bound \ac{(rescaled with the square of the DM abundance)}  from observations by H.E.S.S.\ of the Galactic center (assuming an Einasto profile, brown dot-dashed line) \cite{2016PhRvL.117k1301A} and the expected sensitivity of CTA (also with an Einasto profile, green dashed) \cite{Silverwood:2014yza}, for $\eta\eta \rightarrow W^+W^-$. Right panel: Annihilation fraction of the main channels for two-body final states; $t\bar t$ (blue continuous line), $W^+W^-$ (red dotted), $hh$ (green dot-dashed) and $ZZ$ (black dashed). Other (subdominant) channels are not shown. As in previous plots, the vertical grey lines indicate the range of values of $f$ corresponding to the observed DM abundance at 95\,\% C.L.}
 \label{fig:singletx}
\end{figure}

\section{Conclusions}\label{sec:conclusions}
The amount of evidence for the existence of DM, which comes from astrophysics and cosmology, is overwhelming.
Today, the nature and origin of DM are regarded as one of the biggest problems of contemporary physics. At the same time, a large theoretical effort has been directed towards solving the gauge hierarchy problem. Therefore, the possibility of establishing a link between the two is a tantalizing idea.
This is further supported by the the so-called WIMP miracle. As it is well-known, a WIMP of roughly TeV mass scale can help to explain the inferred DM abundance through the simple freeze-out mechanism.

In spite of the current lack of definite new physics signals at energies of the order of a few TeV, the aforementioned ideas are still widely acknowledged to be excellent reasons to expect a discovery at the LHC in the coming years. Moreover, ongoing direct and indirect detection experiments are also promising windows for the detection of DM particles at the TeV scale, and the sensitivity of these techniques will keep increasing in the near future.

We have worked out a non-minimal CHM containing
a Higgs doublet and three additional scalars: two electrically charged and a neutral one. Depending on
how the SM gauge interactions break the global symmetry, they can either transform as a whole $SU(2)_L$
triplet or as three singlets. Contrary
to the minimal CHM, this setup can explain a large fraction of the observed
DM relic abundance. Moreover, significant improvements with respect to the corresponding elementary extensions of the
scalar sector  are also present. Indeed, if the global symmetries are broken mainly in the fermion sector, our setup
depends on a single parameter ($f$) and the external $\mathbb{Z}_2$ symmetry stabilizing the
DM candidate is \textit{predicted} to be exact also after EW
symmetry  breaking. Were this not the case, the potential would receive sizable contributions
from the gauge sector. These would affect this phenomenological study in several ways:
The relevant observables would not only depend on $f$, but also on $\tilde{c}_1 g_\rho$. In order for the
neutral scalar not to break the $\mathbb{Z}_2$ symmetry by taking a VEV, the condition $\tilde{c}_1 g_\rho^2 \gtrsim -2\pi^2\lambda_H/g^2\sim -7$ should hold. In any case, for $\tilde{c}_1 \sim g_\rho \sim 1$, the bounds on $f$ would be modified only by a small amount.

Assuming that the breaking of $SO(7)$ is driven mainly by the fermion sector, the fraction of the DM abundance that can be accounted for in this framework depends on how the three additional scalars are arranged. In the case in which they form a triplet, the scale $f$ is constrained to be below $\sim4.25$~TeV by H.E.S.S.\ observations of gamma rays from the Galactic center. These would imply that at most $\sim$ 36--46\,\% of the DM abundance can be explained with this model, depending on the shape of the DM radial distribution in the Galactic center. This bound is relatively uncertain, precisely due to our lack of detailed knowledge about the DM profile in the innermost regions of the Galaxy and the modelling of the gamma ray background in that region. Conversely, CMB limits on the DM annihilation cross section are less stringent (though more robust) and allow to account for $\sim$ 80\,\% of the DM abundance with the triplet model. Since the relic density grows (approximately quadratically) with the mass of the DM particle, and the tuning of the EW scale needed to reproduce the correct Higgs mass grows also like $f^2$, there is a linear dependence between tuning and relic density. Therefore,  it is clear that the values of $f$ that would be needed to account for the totality of the DM could be regarded as less natural than those suggested by the current indirect detection constrains. We would like to stress that the mild tuning corresponding to values $f$ giving $\Omega_m h^2 \sim 0.12$ is still acceptable since the Higgs mass is stable under radiative corrections by construction. 

We stress that these results assume a standard thermal history of the Universe. A different thermal history, which in principle is compatible with the model, could help to allow to account for a higher percentage of the DM relic abundance in the triplet case, and would change the upper bound on $f$.

In the case in which the three additional scalars are arranged as three singlets, the neutral one can currently explain the totality of the DM relic abundance. This is simply because in this case the theoretical prediction for the DM annihilation cross section is well below all the current indirect detection upper bounds. In this case, the tuning required to explain the totality of the DM is only increased by a mild factor $\sim 3-10$ with respect to the expectation from naturalness arguments (see for instance Reference~\cite{Panico:2012uw}).

Future observations of the Galactic center from the Cherenkov telescope CTA are expected to improve the sensitivity on the cross section for several DM annihilations channels. However, for DM annihilating into $W^+W^-$ --which is the \gb{common} channel of interest for both of our scenarios-- the analysis of \cite{Silverwood:2014yza} indicates that the sensitivity that will be achieved with CTA is not expected to increase significantly beyond the current H.E.S.S.\ upper bounds in the range of possible DM masses that are relevant for us. 

\gb{However, a forecast of the CTA sensitivity to monochromatic gamma ray lines (produced by DM annihilation into two photons)  \cite{Ovanesyan:2014fwa} indicates that testing most of the relevant range of $f$ for DM in the triplet case should be possible with this channel. Instead, this kind of search is not that useful in the singlet case, since the cross section is much smaller than any current or expected future bound.}

Searches for disappearing tracks performed at the LHC require $f$
to be larger than 650 GeV in the triplet case, while Higgs measurements rise this bound up to $f\sim 800$ GeV in either scenario. Future facilities could improve this bound by almost a factor of 2. Likewise, current direct searches are not constraining, while future experiments would be able to proble all allowed values of $f$.

Clearly, the different searches are rather complementary. Also,
we have set a robust upper limit on the compositeness scale. In generic
CHMs, the latter can be obtained only if (less definite) fine tuning
arguments are advocated. Note also that this bound translates into an upper
limit on the mass, $M$, of the fermionic resonances (roughly speaking, 
$M \lesssim f$). Consequently, a comment on the implications of our
findings for the phenomenology of heavy vector-like fermions is necessary.
In particular, let us focus on top-like resonances, for these are
the ones whose interaction with the SM sector is stronger. These states can be produced in pairs in proton-proton collisions. The
production cross section is mainly driven by QCD interactions, and hence
model independent. Experimental limits on the mass of these resonances
rely only on their branching ratio into the different lighter particles.
Searches performed in the LHC Run I constrain their masses to be
smaller than $\sim 900$ GeV (see for example Reference\,\cite{Aad:2015kqa}). 
More recent analyses~\cite{ATLAS-CONF-2016-101} have pushed this limit just  above
the TeV. The reach of current analyses is still far
from the largest mass allowed by DM experiments. In this respect, our model --and also generic non-minimal CHMs with EW-charged DM candidates--,
favours a hadronic high-energy collider as physics case for a future facility.
On top of that, all current studies consider that the new fermions decay
\textit{only} into SM particles, not into other light scalars expected
in non-minimal CHMs. So, if these setups are to be considered seriously,
and they should, new dedicated searches need to be developed straight
away (see References\,\cite{Serra:2015xfa,Anandakrishnan:2015yfa,Cacciapaglia:2015eqa,
Fan:2015sza, Banerjee:2016wls,Niehoff:2016zso} for works in this direction).

\acknowledgments
{We would like to thank Marco Cirelli, Richard Ruiz, Javi Serra and Marco Taoso for useful discussions. \gb{We also thank Javi Serra and Marco Taoso for comments and suggestions on a draft version of this paper.} The work of G.B. and A.C.\ is funded by the European Union's Horizon 2020 research and innovation programme under the Marie Sk\l{}odowska-Curie grant agreements number 656794 (DEFT) and 659239 (NP4theLHC14), respectively. The work of MC is partially supported by the Spanish MINECO under grant FPA2014-
54459-P and by the Severo Ochoa Excellence Program under grant SEV-2014-0398. G.B.\ thanks the CERN Theoretical Physics Department for hospitality while part of this work was developed.}
\appendix
\section{Representation theory of $SO(7)$ and $G_2$}\label{sec:app1}
Let us define the following $8\times 8$ matrices \cite{Evans:1994np, Chala:2012af}:
\begin{align}
	\gamma_1&=i\sigma_2\otimes i\sigma_2 \otimes i\sigma_2, \quad \gamma_2=\sigma_1\otimes i\sigma_2 \otimes 1, \quad \gamma_3=i\sigma_2\otimes 1\otimes \sigma_1,\quad \gamma_4=-i\sigma_2\otimes 1\otimes \sigma_3,\nonumber\\
	\gamma_5&=1\otimes \sigma_1\otimes i\sigma_2,\qquad \gamma_6=-\sigma_3\otimes i\sigma_2 \otimes 1, \quad \gamma_7=-1\otimes \sigma_3\otimes i\sigma_2
\end{align}
An 8-dimensional representation of $SO(7)$ is then given by the operators
\begin{equation}
	J_{mn}=-J_{mn}=-[\gamma_m,\gamma_n]/4, \qquad m,n=1,\ldots,7. 
\end{equation}
In this paper we consider instead an equivalent representation obtained by rotating $J_{mn}$ (\textit{i.e.} $J_{mn} \rightarrow S^{\dagger}J_{mn}S$) with the following $S$ matrix:
\begin{equation}
	S=\frac{1}{2}\begin{pmatrix}0&1&1&-1&0&0&1&0\\
		1&0&0&0&1&1&0&1\\
		0&1&-1&-1&0&0&-1&0\\
		-1&0&0&0&-1&1&0&1\\
		0&1&-1&1&0&0&1&0\\
		1&0&0&0&-1&1&0&-1\\
		0&1&1&1&0&0&-1&0\\
		-1&0&0&0&1&1&0&-1
	\end{pmatrix}.
\end{equation}
The Lie algebra of $G_2\subset SO(7)$ and the coset space are expanded, respectively, by the following 14 generators, $F_i, M_i$, and the 7 generators, $N_i$~\cite{Gunaydin:1973rs, Chala:2012af}:
\begin{align}
	F_1&=-\frac{i}{2}(J_{24}-J_{51}),\quad M_1=+\frac{i}{\sqrt{12}}(J_{24}+J_{51}-2 J_{73}),\quad N_1=\frac{i}{\sqrt{6}}(J_{24}+J_{51}+J_{73}),\nonumber\\
	F_2&=+\frac{i}{2}(J_{54}-J_{12}),\quad M_2=-\frac{i}{\sqrt{12}}(J_{54}+J_{12}-2 J_{67}),\quad N_2=\frac{i}{\sqrt{6}}(J_{54}+J_{12}+J_{67}),\nonumber\\
	F_3&=-\frac{i}{2}(J_{14}-J_{25}),\quad M_3=+\frac{i}{\sqrt{12}}(J_{14}+J_{25}-2 J_{36}),\quad N_3=\frac{i}{\sqrt{6}}(J_{14}+J_{25}+J_{36}),\nonumber\\
	F_4&=-\frac{i}{2}(J_{16}-J_{43}),\quad M_4=+\frac{i}{\sqrt{12}}(J_{16}+J_{43}-2 J_{72}),\quad N_4=\frac{i}{\sqrt{6}}(J_{16}+J_{43}+J_{72}),\nonumber\\
	F_5&=-\frac{i}{2}(J_{46}-J_{31}),\quad M_5=+\frac{i}{\sqrt{12}}(J_{46}+J_{31}-2 J_{57}),\quad N_5=\frac{i}{\sqrt{6}}(J_{46}+J_{31}+J_{57}),\nonumber\\
	F_6&=-\frac{i}{2}(J_{35}-J_{62}),\quad M_6=+\frac{i}{\sqrt{12}}(J_{35}+J_{62}-2 J_{71}),\quad N_6=\frac{i}{\sqrt{6}}(J_{35}+J_{62}+J_{71}),\nonumber\\
	F_7&=+\frac{i}{2}(J_{65}-J_{23}),\quad M_7=-\frac{i}{\sqrt{12}}(J_{65}+J_{23}-2 J_{47}),\quad N_7=\frac{i}{\sqrt{6}}(J_{65}+J_{23}+J_{47}).
\end{align}
$\lbrace F_1, F_2, F_3\rbrace$ and $\sqrt{3}\,\lbrace M_1, M_2, M_3\rbrace$ span two separate copies of $SU(2)$.
In this particular basis, the vacuum (\textit{i.e.} the vector that it is annihilated only by the generators of $G_2$) adopts the form $\Sigma_0=(0,0,0,0,0,0,0,f)^{T}$.
The $\mathbb{Z}_2$-even spurion $\Delta_q^\alpha$ for the triplet case is given by
\begin{equation}
  \Delta_q^1 = \dfrac{1}{2}\left(\begin{array}{cccccccc}
	  &  & & & & & & 0 \\
	   &      & & &  & & &0 \\
		&     & & &  & & &0 \\
		 \multicolumn{7}{c}{\multirow{7}{*}{\raisebox{90mm}{\scalebox{1.5}{$\mathbf{0}_{7\times 7}$}}}} &0 \\
		 &    & & & & & &0 \\
		   &  & & & & & &-i \\
			& & & & & & &1 \\
			0 & 0 & 0 &0& 0 & -i & 1 & 0 \\
           \end{array}\right)~, \qquad \Delta_q^2 = \dfrac{1}{2}\left(\begin{array}{cccccccc}
	  &  & & & & & & 0 \\
	   &      & & &  & & &0 \\
		&     & & &  & & &0 \\
		 \multicolumn{7}{c}{\multirow{7}{*}{\raisebox{90mm}{\scalebox{1.5}{$\mathbf{0}_{7\times 7}$}}}} &i \\
		 &    & & & & & &1 \\
		   &  & & & & & &0 \\
			& & & & & & &0 \\
			0 & 0 & 0 &i& 1 & 0 & 0 & 0 \\
           \end{array}\right).
\end{equation}
In the singlet case, $\Delta_{q}^1$ is changed by $\Delta_q^{1\ast}$ whereas $\Delta_q^2$ remains unchanged. In this case, the spurion for $\Delta_b$ reads
\begin{equation}
	\Delta_b = \frac{1}{4\sqrt{3}}
	\left(
\begin{array}{cccccccc}
 0 & 0 & -i & 0 & 0 & 0 & 0 & 3 \\
 0 & 0 & 1 & 0 & 0 & 0 & 0 & 3 i \\
 i & -1 & 0 & 0 & 0 & 0 & 0 & 0 \\
 0 & 0 & 0 & 0 & 0 & -i & -1 & 0 \\
 0 & 0 & 0 & 0 & 0 & -1 & i & 0 \\
 0 & 0 & 0 & i & 1 & 0 & 0 & 0 \\
 0 & 0 & 0 & 1 & -i & 0 & 0 & 0 \\
 -3 & -3 i & 0 & 0 & 0 & 0 & 0 & 0 \\
\end{array}
\right).
\end{equation}

 \section{$SU(2)_L\times U(1)_Y$ quantum numbers of the pNGBs}\label{sec:app2}
 To recognize which combination of pNGBs spans  the $\mathbf{2}_{1/2}$ or the $\mathbf{3}_0$ of the EW group it is useful to remember that, if the broken generators $X^a$ transform as
 \begin{align}
	 \mathrm{exp}(-\alpha^i Y_i) \, X^a\, \mathrm{exp}(\alpha^j Y_j)=R_{ab}\, X^a
 \end{align}
 under an element $h=\mathrm{exp}(\alpha_i Y^i)$ of the unbroken group $G_2$, the pNGBs accompanying them inside  $U=\mathrm{exp}\left(i\Pi^a N^a/f\right)$ transform with the transposed matrix, \textit{i.e.},
 \begin{align}
	 \Pi^a\to R_{ab}^T \,\Pi^b~.
 \end{align}
 For simplicity, let us focus first on the triplet case. If we define
 \begin{eqnarray}
 	N^{\pm}\equiv \frac{N^1\pm iN^2}{\sqrt{2}}, \qquad N^0\equiv -N^3,\quad \mathrm{and}\quad 	N_{\Phi}=\begin{pmatrix}N^+\\ N^0 \\ -N^-\end{pmatrix}, 
 \end{eqnarray}
 and use their commutations relations, we get
 \begin{eqnarray}
 	\sqrt{3}[M^i,N_{\Phi}]=-t^i N_{\Phi}, \qquad [F^3,T_{\Phi}]=-0 N_{\Phi},
 	\label{eq:ap}
 \end{eqnarray}
 where $t^i,~i=1,2,3,$ are the three-dimensional $SU(2)$ representation given in (\ref{eq:gen1}).
 Therefore, 
 \begin{align}
 	e^{-i\alpha_j \sqrt{3}M^j} N_{\Phi} e^{i\alpha_k \sqrt{3}M^k} & = N_{\Phi}-i\alpha_j\sqrt{3}[M^j,N_{\Phi}]+\ldots \nonumber\\
 	& =  (1+i\alpha_jt^j)N_{\Phi}+\ldots =e^{i\alpha_j t^j}N_{\Phi}
 \end{align}
 and 
 \begin{eqnarray}
	 \Phi^{\ast}\to \left(e^{i\alpha_j t^j}\right)^T\Phi^{\ast} 	\Rightarrow \Phi\to e^{-i\alpha_j t_j}\Phi~,
 \end{eqnarray}
 where we have defined
 		\begin{eqnarray}
			\Phi=\begin{pmatrix}\kappa^{+}\\-\eta \\ -\kappa^-\end{pmatrix},\quad \mathrm{and}\quad	\kappa^{\pm}=\frac{\kappa^1\pm \kappa^2}{\sqrt{2}}.
 		\end{eqnarray}
		This means that
	 $\Phi$ transforms properly as a hyperchargeless $SU(2)$ triplet. Analogously, if we define
 	\begin{align}
		N_{H}=\frac{1}{\sqrt{2}}\begin{pmatrix}N^3-iN^4\\ N^6+i N^7 \end{pmatrix}
 	\end{align}
 	and use the commutation relations we get
 \begin{eqnarray}
 	\sqrt{3}[M^i,N_{H}]=-\frac{1}{2}\sigma^i N_{H}, \qquad [F^3,N_{H}]=-\frac{1}{2} N_{H}~,
 	\label{eq:ap2}
 \end{eqnarray}
 which implies that 
 	\begin{eqnarray}
		H=\frac{1}{\sqrt{2}}\begin{pmatrix}h^1-ih^2\\ h^3+ih^4 \end{pmatrix}
 	\end{eqnarray}
  transforms as an $SU(2)$ doublet with $Y=1/2$ hypercharge. In the singlet case, this combination can be taken 
 	\begin{eqnarray}
		H=\frac{1}{\sqrt{2}}\begin{pmatrix}-h^1+ih^2\\ h^3-ih^4 \end{pmatrix}~.
 	\end{eqnarray}
 \section{The case of composite leptons}\label{sec:app3}
An interesting possibility that has been explored recently is that leptons could play a role in EWSB when they transform in non minimal irreps of the global group,  see \textit{e.g.} References\,\cite{Carmona:2014iwa, Carmona:2015ena, Carmona:2016mjr}. (By non-minimal irreps we mean that they can provide more than one independent invariant under the unbroken group at leading order in the spurion expansion, like \textit{e.g.} the $\mathbf{14}$ in $SO(5)/SO(4)$ or the $\mathbf{35}$ in $SO(7)/G_2$.)
The rationale is that, when the quark sector transforms in smaller representations of the Goldstone symmetry (like the spinorial, the fundamental, the adjoint, ...), even a moderate degree of compositeness in one of the lepton chiralities can have a sizable impact in the Higgs potential. This is due to the fact that the leading lepton contribution to the Higgs quartic coupling scales in this case with $|\lambda_{\ell}|^2/g_{\ast}^2$, whereas the top one goes with  $|\lambda_{q}|^4/g_{\ast}^4$, $|\lambda_q|^2|\lambda_{t}|^2/g_{\ast}^4$ or $|\lambda_{t}|^4/g_{\ast}^4$. Therefore, a relatively smaller value of $\lambda_{\ell}/g_{\ast}$ arising from the charged lepton sector can provide a comparable effect to the one coming from the top quark. 

Moreover, the fact that all different lepton generations could be partially composite, could enhance the lepton contribution by a factor $N_{\rm gen}\sim 3$,  compensating the color factor $N_c=3$ present in the top case. Indeed, the recent hints of violation of lepton flavor universality observed by LHCb and CMS in $R_K$ and $R_K^{\ast}$  \cite{Aaij:2014ora, rkstar} seem to provide a further motivation to these scenarios, as discussed \textit{e.g.} in \cite{Carmona:2015ena}.

In what follows, we will briefly discuss how a similar setup works in the case of $SO(7)/G_2$ and its impact on  DM. We assume that $\mathcal{O}_q^j$ and $\mathcal{O}_u^j$ transform in the $\mathbf{8}_{2/3}$ and the $\mathbf{1}_{2/3}$ of $SO(7)\times U(1)_X$ , respectively, whereas the composite operators mixing with the left-handed lepton doublets and the right-handed charged singlets,  $\mathcal{O}_{L}^j$ and $\mathcal{O}_{\ell}^j$, transform respectively in the $\mathbf{1}_{-1}$ and $\mathbf{35}_{-1}$ of the same group. Then, the scalar potential can be written as

\begin{align}
	V(\Pi)\approx m_{\ast}^2 f^2 \frac{1}{16\pi^2} \left[N_c \left(\frac{|\lambda_q|}{g_{\ast}}\right)^2c_1V_{1}(\Pi) +\sum_{j=1}^3\left(\frac{|\lambda_{\ell}|_{jj}}{g_{\ast}}\right)^2\left[c_{2,j}V_{2}(\Pi)+c_{3,j} V_3(\Pi)\right]\right],
	\label{eq:pot2}
\end{align}
\begin{table}[t]
	\begin{center}
	\begin{adjustbox}{width=\textwidth}
		\begin{tabular}{lccc}
			\toprule
			Parameter&$\mu_{H}^2$&$\lambda_{\Phi}$&$\lambda_{H\Phi}$\\
			\midrule
			Value&$-v^2\lambda_H$&$\frac{16}{9}\lambda_{H}\left(1-\frac{2}{3}\frac{v^2}{f^2}\right)-\frac{2}{3}\frac{\mu_{\Phi}^2}{f^2}$&$-\frac{4}{3}\lambda_{H}\left(1-\frac{11}{9}\frac{v^2}{f^2}\right)-\frac{1}{3}\frac{\mu_{\Phi}^2}{f^2}$\\
						\bottomrule
		\end{tabular}
		\end{adjustbox}
		\caption{Values of the different parameters of the renormalizable scalar potential as a function of $f$, to order $\mathcal{O}(v^2/f^2)$. $v\sim 246$ GeV and $\lambda_H\sim 0.13$ stand for the \mc{SM} EW VEV and the Higgs quartic coupling, respectively.}
		\label{tab:pot2}
	\end{center}

\end{table}
with 
\begin{align}
	\sum_{\alpha}	\left|(\Delta_{qD}^{\alpha})_{8}\right|^2 \sim V_1(\Pi)&=\frac{|H|^2}{\hat{\Pi}^2}\sin^2\left(\frac{\hat{\Pi}}{f}\right), \\
	\left|(\Delta_{\ell D})_{88}\right|^2 \sim V_2(\Pi)&=\frac{8}{147}\frac{1}{\hat{\Pi}^4}\sin^4\left(\frac{\hat{\Pi}}{f}\right)\left(3|H|^2-2|\Phi|^2\right)^2,\\
	\sum_{i=1}^7	\left|(\Delta_{\ell D})_{i8}\right|^2 \sim V_3(\Pi)&=\frac{1}{21}\frac{1}{\hat{\Pi}^4}\sin^2\left(\frac{\hat{\Pi}}{f}\right)\left[2\left(3|H|^2-2|\Phi|^2\right)^2\cos^2\left(\frac{\hat{\Pi}}{f}\right)+49|H|^2|\Phi|^2\right],
\end{align}
where we have defined the dressed spurions $\Delta_{qD}^{\alpha}=U^{-1}\Delta_q^{\alpha},~\alpha=1,2$, and $\Delta_{\ell D}=U^{-1}\Delta_{\ell}U$ as usual, with\,\footnote{We are thinking of the triplet case. In the singlet case one has to change $\Delta_q^1$ by $\Delta_q^{1\ast}$, with the rest of the spurions remaining the same.}
\begin{align}
	\Delta_q^1=\frac{1}{\sqrt{2}}\begin{pmatrix}0\\ 0\\0\\0\\0\\-i\\1\\0\end{pmatrix}, \quad \Delta_q^2=\frac{1}{\sqrt{2}}\begin{pmatrix}0\\ 0\\0\\i\\1\\0\\0\\0\end{pmatrix},\quad 
		\Delta_{\ell}=\frac{1}{2}\sqrt{\frac{3}{7}}\,\mathrm{diag}\begin{pmatrix}4/3\\4/3\\ 4/3\\-1\\-1\\-1\\-1\\0\end{pmatrix}.
\end{align}
The parameters  $c_1,c_{2,j}$ and $c_{3,j}$, with $j=1,2,3,$ running over the three lepton generations, that appear in the scalar potential are order one dimensionless numbers. Note, however, that $c_{2,j}$ and $c_{3,j}$ always enter in the same  linear combination. So, effectively, we are left with  only three independent unknowns   (the coefficients of $V_1(\Pi), V_2(\Pi)$ and $V_3(\Pi)$) which can be traded at the renormalizable level by  the Higgs VEV $v$, the Higgs quartic $\lambda_H$ and the mass parameter of the  scalar triplet $\mu_{\Phi}^2$, see Table~\ref{tab:pot2} and Equation\,\ref{eq:ren}. The mass of the triplet in the EW phase is given by
\begin{align}
m_{\Phi}^2=\mu_{\Phi}^2+\lambda_{H\Phi}v^2=\mu_{\Phi}^2\left[1-\frac{v^2}{3f^2}+\mathcal{O}\left(\frac{v^4}{f^4}\right)\right]-\frac{4}{3}\lambda_{H}v^2\left[1-\frac{11}{9}\frac{v^2}{f^2}+\mathcal{O}\left(\frac{v^4}{f^4}\right)\right]~.
\end{align}
Since $\mu_{\Phi}^2\sim f^2\gg \lambda_H v^2$, the triplet  does not take a VEV provided the underlying UV dynamics allows for a positive   $\mu_{\Phi}^2$ (the same holds for the singlet if we weakly gauge the other $SU(2)$ as discussed in Section\,\ref{sec:singlet}, since the main contribution to the potential is still given by Equation\,\ref{eq:ren}). The main difference with respect to the scenarios explored before is that the relationship between $\mu_{\Phi}$ and $f$ is in principle not known. However, the same phenomenological study could be done having as an extra variable the ratio $\mu_{\Phi}/f$, what we leave for a future work.

\bibliographystyle{JHEP}
\bibliography{references}{}
\end{document}